\documentclass[pre,aps,10pt,amsfonts,floatfix,superscriptaddress,longbibliography,twocolumn]{revtex4-2}
\usepackage{amsmath, amssymb, bm, physics, mathbbol}
\usepackage{epstopdf}
\usepackage{mciteplus}
\usepackage{bbding}
\usepackage[utf8]{inputenc}
\usepackage[titletoc]{appendix}
\usepackage{braket}
\usepackage{float, subfigure} 
\usepackage[graphicx]{realboxes}
\usepackage{graphicx,color}
\usepackage{hyperref}
\usepackage{soul}
\hypersetup{colorlinks=true, linkcolor=red, urlcolor=magenta}

\begin{document}

\title{Reducing dynamical fluctuations and enforcing self-averaging by opening many-body quantum systems}

\author{Isa\'ias Vallejo-Fabila}
\affiliation{Department of Physics, University of Connecticut, Storrs, Connecticut 06269, USA}
\author{Adway Kumar Das}
\affiliation{Department of Physics, University of Connecticut, Storrs, Connecticut 06269, USA}
\affiliation{Department of Physical Sciences, Indian Institute of Science Education and Research Kolkata, Mohanpur 741246, India}
\author{David A. Zarate-Herrada}
\affiliation{Institut of Physics, Benem{\'e}rita Universidad Aut{\'o}noma de Puebla, Puebla, 72570, Mexico}
\author{Apollonas S. Matsoukas-Roubeas}
\affiliation{Department of Physics and Materials Science, University of Luxembourg, L-1511 Luxembourg, Luxembourg}
\author{E. Jonathan Torres-Herrera}
\affiliation{Institut of Physics, Benem{\'e}rita Universidad Aut{\'o}noma de Puebla, Puebla, 72570, Mexico}
\author{Lea F. Santos}
\affiliation{Department of Physics, University of Connecticut, Storrs, Connecticut 06269, USA}



\begin{abstract}
We investigate how the dynamical fluctuations of many-body quantum systems out of equilibrium can be mitigated when they are opened to a dephasing environment. We consider the survival probability (spectral form factor with a filter) evolving under different kinds of random matrices and under a spin-1/2 model with weak and strong disorder. In isolated many-body quantum systems, the survival probability is non-self-averaging at any timescale, that is, the relative variance of its fluctuations does not decrease with system size. By opening the system, we find that the fluctuations are always reduced, but self-averaging can only be ensured away from critical points. Self-averaging is achieved for the long-time dynamics of full random matrices, power-law banded random matrices deep in the delocalized phase, and the Rosenzweig-Porter ensemble in all the phases except at the 
localization transition point. 
For the spin model, the survival probability becomes self-averaging only in the chaotic regime provided the initial states are in the middle of the spectrum. 
Overall, a strongly non-self-averaging survival probability in open systems is an indicator of criticality.
\end{abstract}

\maketitle


\section{Introduction}
\label{Sec: Introduction}

The effects of the environment on a quantum system are often considered to be detrimental in the context of quantum technologies. External interactions usually result in the rapid loss of quantum coherence, which hinders the realization of quantum information processing and has motivated the development of methods to reduce decoherence. However, the influence of the environment can also be beneficial. For example, signatures of quantum chaos can be enhanced by certain kinds of non-Hermitian evolution facilitating their study \cite{Xu2019,Cornelius2022,Matsoukas2023}. Dynamical fluctuations can also be decreased by slightly opening a system to a dephasing environment~\cite{Tameshtit1992,delcampo19,Xu2021,Xu2021,Matsoukas2023,Matsoukas2023quantum}, a strategy that has been used to achieve self-averaging for the spectral form factor of random matrices~\cite{Matsoukas2023pra}. The present work explores the effects of a dephasing environment on the dynamical fluctuations of experimental models. 

An observable $\mathcal{O}$ is self-averaging when its relative variance, that is, the ratio between its variance and the square of its mean~\cite{Wiseman1995,Aharony1996,Wiseman1998,Castellani2005,Malakis2006,Roy2006,Monthus2006,Efrat2014,Lobejko2018},
\begin{equation}
R_{\mathcal{O}}(t)=\frac{\sigma_{\mathcal{O}}^{2}(t)}{\braket{\mathcal{O}(t)}^{2}}=\frac{\braket{\mathcal{O}^{2}(t)}-\braket{\mathcal{O}(t)}^{2}}{\braket{\mathcal{O}(t)}^{2}},
\label{Eqn: RV(t)}  
\end{equation}
goes to zero as the system size increases, where $\braket{\cdot}$ indicates  average over an ensemble. When this happens, the observable does not fluctuate in the thermodynamic limit~\cite{Gredeskul1985}.
The presence of self-averaging is important because it implies that the number of samples used in experiments and in numerical analysis can be decreased as the system size $L$ increases, and theoretical models can be built to describe the physical properties of $\mathcal{O}$ using finite samples.

Lack of self-averaging is usually associated with the critical point of disordered systems at equilibrium~\cite{Wiseman1995,Aharony1996,Wiseman1998,Castellani2005,Malakis2006,Roy2006,Monthus2006,Efrat2014,MullerARXIV,Serbyn2017,Pastur2014PRL,Milchev1986,Bouchaud1990,AkimotoPRL2016,Russian2017,AkimotoPRE2018,Wreszinski2004}. In studies of many-body localization, lack of self-averaging hinders scaling analysis~\cite{Solorzano2021}, because in addition to having to deal with a Hilbert space that grows exponentially with $L$, the number of samples cannot be reduced as the system size increases.

Self-averaging properties are also studied in systems out of equilibrium~\cite{Ithier2017,Lobejko2018,Mukherjee2018,Richter2020,Schiulaz2020,Torres2020PRE,Torres2020}, as indicated with the time dependence of $R_{\mathcal{O}}(t)$ in Eq.~(\ref{Eqn: RV(t)}). Lack of self-averaging has been observed for large time intervals even in the chaotic regime, as shown in the spectral form factor \cite{Argaman1993b,Eckhardt1995,Prange1997,Braun2015}. The spectral form factor is the Fourier transform of the two-point correlation function of the energy spectrum~\cite{MehtaBook},
\begin{align}
	\label{eq_SFF}
	\mathrm{SFF}(t) = \dfrac{1}{N^2} \left\langle \sum_{n, m}e^{- i(E_n - E_m)t} \right\rangle,
\end{align}
where $\hbar =1$, $N$ is the dimension of the Hilbert space, and $E_n$'s are the eigenvalues of the system Hamiltonian with spectral decomposition $H = \sum_nE_n|E_n\rangle \langle E_n\rangle$. In chaotic systems, $\mathrm{SFF}(t)$ presents a slope-dip-ramp-plateau structure~\cite{Leviandier1986,Pique1987,Guhr1990,Hartmann1991,Alhassid1992,Lombardi1993,Michaille1999,Leyvraz2013,Torres2017Philo,Torres2017,Torres2018,Schiulaz2019,Das2024,Lerma2019,Santos2020,Campo2018a,Xu2021,Lezama2021,Das2022a,Das2022b,Das2023b,shir2023range,Cotler2017,Cotler2017GUE,Suntajs2020,Dag2023,dong2024measuring}, known as correlation hole~\cite{Leviandier1986,Pique1987,Guhr1990,Hartmann1991,Alhassid1992,Lombardi1993,Michaille1999,Leyvraz2013,Torres2017Philo,Torres2017,Campo2018a,Torres2018,Schiulaz2019,Das2024,Lerma2019,Santos2020,Xu2021,Lezama2021,Das2022a,Das2022b,Das2023b,shir2023range}, that detects short- and long-range level correlations similar to those in full random matrices. The absence of self-averaging for the spectral form factor implies that ensemble averages are necessary for revealing the correlation hole.

The spectral form factor can be interpreted as the average survival probability,  
\begin{eqnarray}
	\label{eq_SP_def}
		\left\langle S_P(t) \right\rangle &=&\left\langle | \langle \Psi(0) |\Psi(t) \rangle |^2  \right\rangle  \\
  &=&  \left\langle \sum_{n,m}  |c_n^{(0)}|^2 |c_m^{(0)}|^2 e^{- i(E_n - E_m)t}  \right\rangle, \nonumber
\end{eqnarray}
 of an initial state $|\Psi(0)\rangle$, where the coefficients $c_n^{(0)} = \langle E_n|\Psi(0)\rangle$ play the role of a filter for $\mathrm{SFF}(t)$.
When the initial state is a coherent Gibbs state,   $|c_n^{(0)}|^2  =e^{-\beta E_n}/Z(\beta)$ are the Boltzmann factors, $Z(\beta)=\sum_{n=1}^N e^{-\beta E_n}$, and $\beta$ is the inverse temperature~\cite{Campo2018a,Campo2018}.  We recover Eq.~(\ref{eq_SFF}) when $\beta=0$. To facilitate the connection with experiments, it is also common to investigate the survival probability of initial states defined via quench dynamics~\cite{Torres2017Philo,Torres2017,Torres2018,Schiulaz2019}.

In Ref.~\cite{Schiulaz2020}, it was shown numerically and analytically that the survival probability evolving under full random matrices is non-self-averaging at any timescale. The same happens for spin models quenched in the chaotic regime~\cite{Schiulaz2020,Torres2020PRE} and away from chaos~\cite{Torres2020}. A way to solve the problem of lack of self-averaging for the long-time dynamics of the survival probability of coherent Gibbs states was achieved in Ref.~\cite{Matsoukas2023pra} by opening the system to a dephasing environment. Using random matrices, it was shown that the fluctuations in the values of the survival probability are reduced and self-averaging is ensured after the saturation of the dynamics~\cite{Matsoukas2023pra}.
It was also demonstrated that the use of averages is equivalent to making the time evolution nonunitary, effectively opening the system.

In the present work, we study the self-averaging properties of the survival probability at different timescales in  open quantum systems subject to energy dephasing. We consider systems prepared in initial states given by coherent Gibbs states and from quench dynamics, which are evolved under full random matrices, power-law banded random matrices (PBRM), Rosenzweig-Porter ensembles (RPE) of random matrices, and disordered spin-1/2 models. The PBRM model, RPE, and spin model are explored for different values of a control parameter that moves these systems from a delocalized to a localized phase.  We provide analytical expressions for the relative variance of the survival probability at short and long times, which help in understanding our numerical results. 

We find that, even though the fluctuations of the survival probability are reduced for all four models at all timescales, self-averaging is not always guaranteed. Self-averaging is enforced in open systems characterized by full random matrices, by the RPE away from the delocalization-localization critical point, and by PBRM deep in the delocalized phase. In the case of the spin model, not only the strength of the disorder but also the choice of the initial states can prevent self-averaging. In this case, the relative variance of $S_P(t)$ only decreases with system size deep in the chaotic regime and for initial states in the middle of the spectrum.

The paper is organized as follows. Section~\ref{Sec: General definitions} provides the expression of the survival probability in open systems for Gibbs states and initial states in quench dynamics. In Sec.~\ref{Sec:Ranalytical}, we derive analytical expressions for the relative variance of the survival probability at short and long times.
In Sec.~\ref{Sec: Models}, we analyze the survival probability for full random matrices, where analytical results can be obtained. Next, we consider PBRM and the RPE, which are closer to physical models and allow for investigating delocalized and localized regimes. In Sec.~\ref{Sec: Physical models}, we show the relative variance of the survival probability evolving under a disordered spin-1/2 model for different disorder strengths. Conclusions are given in Sec.~\ref{Sec: Conclusions}.


\section{Survival Probability in an Open system}
\label{Sec: General definitions}

The Lindblad master equation,
\begin{equation}
\dot{\rho}(t)=-i [H,\rho(t)]+\sum_{k}\gamma_{k}\left(L_{k}\rho L_{k}^{\dag}-\frac{1}{2}\lbrace L_{k}L_{k}^{\dag},\rho\rbrace\right),
\label{Eqn: Master equation. General}  
\end{equation}
describes the Markovian dynamics of an open system, where $\rho(t)$ is the evolved density matrix, $\hbar =1$, $H$ is the Hamiltonian of the isolated quantum system, $L_{k}$ is an arbitrary operator, and $\gamma_k\geq 0$. We consider energy dephasing processes with $k=1$, $\gamma_{1}=2\kappa$, and $L_{1}=L_{1}^{\dag}=H$, for which the Lindblad master equation is written as \cite{Tameshtit1992,delcampo19,Xu2019,Xu2021} 
\begin{equation}
\dot{\rho}(t)=-\frac{i}{\hbar}[H,\rho(t)]-\kappa [H,[H,\rho(t)]],
\label{Eqn: Master equation. Energy dephasing}  
\end{equation}
and $\kappa$ is the dephasing strength dependent on the amplitude of the external couplings and the properties of the environment. This evolution is unital (i.e., the maximally mixed state is invariant at all times) 
and gives rise to a monotonic decay of the purity, $\Tr[\rho(t)^2]$, thus making the time-evolving state increasingly mixed as time passes. 

When the initial is pure, $\rho(0)=|\Psi(0)\rangle\langle\Psi(0)|$, as considered here, the 
survival probability for an open system takes the form \cite{Xu2021,Cornelius2022}
\begin{equation}
S_{P}(t)=\langle\Psi(0)|\rho(t)|\Psi(0)\rangle = \Tr[\rho(0) \rho(t)],
\label{Eqn: Sp. Definition}  
\end{equation}
which is the probability that $\rho(t)$ agrees the initial state $|\Psi(0)\rangle\langle\Psi(0)|$. The equation above is equivalent to the average of the survival probability $| \langle \Psi(0) |\Psi(t) \rangle |^2$ over an ensemble of pure states determined by the density matrix $\rho(t)$ \cite{Jozsa1994,Carrera2022}. [Notice that Eq.~(\ref{Eqn: Sp. Definition}) needs to be modified when both $\rho(0)$ and $\rho(t)$ are mixed states.] 

The solution of Eq.~(\ref{Eqn: Master equation. Energy dephasing}), taking Eq.~(\ref{Eqn: Sp. Definition}) into account, is \cite{Tameshtit1992,delcampo19,Xu2019}
\begin{equation}
\rho(t)=\sum_{n,m=1}^{N}e^{-i\omega_{nm}t-\kappa\omega_{nm}^{2}t}\rho_{nm}(0)\ket{E_{n}}\bra{E_{m}},
\label{Eqn: Density matrix}  
\end{equation}
where 
$\omega_{nm}=E_{n}-E_{m}$ 
and
$\rho_{nm}(0) =c_{n}^{(0)}c_{m}^{(0)*}$.
This means that the survival probability is
\begin{equation}
S_{P}(t)=\sum_{n,m=1}^{N}|c_{n}^{(0)}|^{2}|c_{m}^{(0)}|^{2}e^{-i\omega_{nm}t-\kappa\omega_{nm}^{2}t}.
\label{Eqn: Sp. Open system}  
\end{equation}

We analyze three possibilities for the initial states:

(i) A coherent Gibbs state at infinite temperature, $\beta=0$, in which case the average survival probability coincides with the spectral form factor,
\begin{equation}
       \left\langle S_{P}^{\beta=0} (t) \right\rangle= \left\langle \frac{1}{N^{2}}\sum_{n,m=1}^{N}e^{-i\omega_{nm}t-\kappa\omega_{nm}^{2}t} \right\rangle .
        \label{Eqn: SFF. Beta zero}
    \end{equation}
The average $\braket{\cdot}$ is performed over $10^4$ disorder realizations. The energy of this initial state, $E^{(0)} =\langle \Psi(0)|H| \Psi(0) \rangle = \sum|c_n^{(0)}|^2 E_n $, is at the middle of the spectrum.

(ii) A coherent Gibbs state at finite temperature, 
    \begin{equation}
        \left\langle S_{P}^{\beta \neq0} (t) \right\rangle = \left\langle \frac{1}{Z^{2}}\sum_{n,m=1}^{N}e^{-\beta(E_{n}+E_{m})}e^{-i\omega_{nm}t-\kappa\omega_{nm}^{2}t} \right\rangle.
        \label{Eqn: SFF. Beta nonzero}
    \end{equation} 
The average is also performed over $10^4$ disorder realizations.  Our analysis is done for $\beta=0.1$. This initial state involves a coherent superposition that is predominantly composed of low-energy eigenstates associated with the largest values of $|c_n^{(0)}|^2 = e^{-\beta E_n}/Z(\beta)$. This means that this initial state has energy $E^{(0)}$ very close to the lower edge of the spectrum.

(iii) An initial state obtained by quenching a given initial Hamiltonian $H_{0}$  onto a final Hamiltonian $H = H_0 + V$, in which case the post-quench initial state is an eigenstate of $H_0$ and the average survival probability is
\begin{equation}
\left\langle S_{P}^{qch}(t) \right\rangle = \left\langle  \sum_{n,m=1}^{N} |c_n^{(0)}|^2 |c_m^{(0)}|^2  e^{-i\omega_{nm}t-\kappa\omega_{nm}^{2}t} \right\rangle .
\end{equation} 
The average is performed over $n_d \times n_i =10^4$ 
data distributed between $n_d$ disorder realizations and $n_i$ 
initial states. For $L=8$, $10$, $12$, $n_d =1000$, for  $L=14$, $n_d = 500$, and for  $L=16$, $n_d = 200$. We explore the quench dynamics for initial states with energy in the middle and at the edge of the spectrum.

\begin{figure}[t]
\centering
\includegraphics[scale=.75]{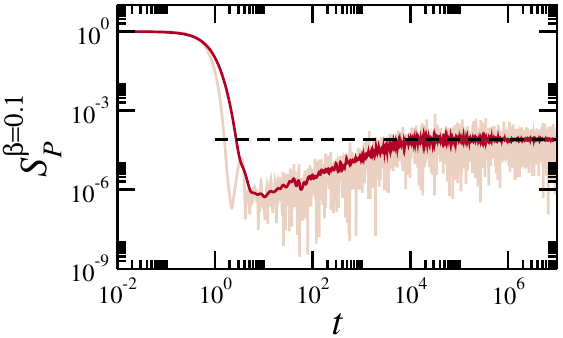}
\caption{The dynamical fluctuations of the survival probability are decreased by opening the system. The figure shows the entire evolution of the survival probability for a Gibbs initial state with $\beta=0.1$ and a single realization of a chaotic spin-1/2 model. 
The light curve represents the isolated model ($\kappa=0.0$) and the dark red curve is for the open system ($\kappa=0.05$). The horizontal dashed line gives the saturation value of the dynamics, which coincides with IPR$_0$ [Eq.~(\ref{Eq:IPRPsi0})]. Time in units of the coupling parameter of the model.}
\label{Fig: Fig01. Reduction of fluctuations}
\end{figure}

The temporal fluctuations of the values of the survival probability are reduced by opening the system~\cite{Tameshtit1992,Xu2021,Matsoukas2023pra}. This is illustrated in Fig.~\ref{Fig: Fig01. Reduction of fluctuations}, where we consider a single coherent Gibbs state with $\beta=0.1$ as the initial state and a single disorder realization of a chaotic spin model and compare the evolution of $S_{P}^{\beta=0.1}(t)$ for the isolated system (light curve) with the evolution of $S_{P}^{\beta=0.1}(t)$ for the open system undergoing energy dephasing (red curve). Since the model is chaotic, the survival probability presents the typical slope-dip-ramp-plateau structure, that is, the correlation hole below the horizontal dashed line that indicates the saturation of the dynamics (plateau). The fluctuations are significantly reduced by opening the system, especially during the ramp toward the plateau and after saturation, which should facilitate the experimental detection on the correlation hole~\cite{Das2024}. 

In the open system, the effects of averages are achieved through the environment. In both cases, that of $S_P(t)$ evolving under an open system and that of the survival probability evolved in an isolated system and averaged over an ensemble, $\langle S_P(t) \rangle$, the evolution is in effect non-unitary~and described by a mix-unitary quantum channel~\cite{Matsoukas2023pra}. 

The question addressed in this work goes beyond the reduction of fluctuations achieved with Eq.~(\ref{Eqn: Sp. Open system}). We want to know whether, by opening the system, the relative fluctuations, determined by $R_{S_P}(t)$ in Eq.~(\ref{Eqn: RV(t)}), decrease with system size, thus ensuring self-averaging.

\section{Relative Variance of the Survival Probability}
\label{Sec:Ranalytical}

Analytical expressions for the relative variance of the survival probability at short and long times for the open system can be obtained as follows.

\subsection{Short times}
\label{SubSecApp: Short times}

Let us start by Taylor expanding $S_P(t)$ for $t \rightarrow 0$. The odd powers $\mu$ of $\omega_{nm}$ cancel, because $\sum_{n,m=1}^N (E_n-E_m)^\mu =0$. Since $\kappa, t \ll 1$, we keep only the terms proportional to $\kappa^{a}t^{b}$ with $a+b\leq4$ and obtain 
\begin{equation*}
\langle S_P(t\to0) \rangle \approx 1 -
\langle d_2 \rangle \left(\kappa t +\frac{t^{2}}{2} \right) + \langle d_4 \rangle  \left(\frac{\kappa^{2}t^{2}}{2}+\frac{\kappa t^{3}}{2}+\frac{t^{4}}{24} \right) ,
\label{Eqn: W(t to 0)}
\end{equation*}
where
\begin{eqnarray}
d_2 &=& \sum_{n,m=1}^{N} |c_n^{(0)}|^2 |c_m^{(0)}|^2 \omega_{nm}^{2} , \nonumber \\
&=& 2 \left[ \sum_{n}^{N} |c_n^{(0)}|^2 E_n^2 - \left( \sum_{n}^{N} |c_n^{(0)}|^2 E_n \right)^2 \right] = 2 \Gamma^2, \,\,\,\,\,\,\,\,\,
\label{Eq:Gamma}
\end{eqnarray}
$\Gamma$ is the width of the energy distribution of the initial state, and
\begin{eqnarray}
d_4 &=& \sum_{n,m=1}^{N} |c_n^{(0)}|^2 |c_m^{(0)}|^2 \omega_{nm}^{4} . \nonumber
\end{eqnarray}
For the average of the squared survival probability, we get
\begin{eqnarray}
\langle S_P(t\to0)^2 \rangle &\approx& 1 - 2
\langle d_2 \rangle \left(\kappa t +\frac{t^{2}}{2} \right) + 
\langle d_2^2 \rangle  \left(\kappa t +\frac{t^{2}}{2} \right)^2 \nonumber \\ 
&+& 2 
\langle d_4 \rangle  \left(\frac{\kappa^{2}t^{2}}{2}+\frac{\kappa t^{3}}{2}+\frac{t^{4}}{24} \right) . \nonumber
\end{eqnarray}
This implies that the relative variance at short times is approximately
\begin{equation}
R_{S_P}(t \rightarrow 0 ) \approx \sigma_{d_2}^2 \left( \kappa^{2}t^{2}  + \kappa t^{3} + \frac{t^4}{4}
\right) ,
\label{Eqn:RVshortTime}  
\end{equation}
where the variance $\sigma_{d_2}^2 = 4 \sigma_{\Gamma^2}^2$. The width $\Gamma$ of the energy distribution of the initial state $|\Psi(0)\rangle$ depends on the number of states directly coupled to it. As the system size grows, the range of values of this number also increases, so we expect $\sigma_{\Gamma^2}$ to grow, which justifies the lack of self-averaging for the survival probability at very short times even after opening the system.

\subsection{Long times}
\label{SubSecApp: Long times}

In the limit $t\to\infty$, the terms in the sum for the survival probability in Eq.~(\ref{Eqn: Sp. Open system}) are zero, unless $n=m$, so the infinite-time average for $\kappa \neq 0$ or for $\kappa = 0$ gives
\begin{equation}
\overline{S_P}=\sum_{n=1}^{N}|c_n^{(0)}|^4 =\text{IPR}_0  .
\label{Eq:IPRPsi0}
\end{equation}
This saturation value corresponds to the ``plateau'', as shown in Fig.~\ref{Fig: Fig01. Reduction of fluctuations}, and is also referred to as the inverse participation ratio (IPR) of the initial state written in the energy eigenbasis.

Similarly, for $\kappa \neq 0$, the mean of the square of the survival probability,
\begin{eqnarray}
\overline{S_P^2}= &&\overline{\sum_{n,m=1}^{N}|c_{n}^{(0)}|^{2}|c_{m}^{(0)}|^{2}e^{-i\omega_{nm}t-\kappa\omega_{nm}^{2}t} } \nonumber \\
&& \times \overline{\sum_{p,q=1}^{N}|c_{p}^{(0)}|^{2}|c_{q}^{(0)}|^{2}e^{-i\omega_{pq}t-\kappa\omega_{pq}^{2}t},
} 
\label{Eq:SPsquare}
\end{eqnarray}
is nonzero for $n=m$ and $p=q$.
According to Eq.~(\ref{Eq:IPRPsi0}), the infinite-time average of the relative variance of $S_P(t)$ is then given by the relative variance of the IPR of the initial state,
\begin{equation}
\overline{R_{S_P}^{\kappa \neq 0} } = \frac{ \sigma_{\text{IPR}_0}^{2} }{\braket{\text{IPR}_0}^{2}}.
\label{Eqn: Relative variance of W --- long times}  
\end{equation}
Therefore, the scaling of $\overline{R_{S_P}^{\kappa \neq 0} }$ with the system size of open systems depends on the model and the initial states. 

Equation~(\ref{Eqn: Relative variance of W --- long times}) is a crucial result of this paper. It indicates that the long-time analysis of the self-averaging properties of the survival probability in open systems boils down to the analysis of the self-averaging behavior of the initial state, that is, of IPR$_0$. This means that the structures of the eigenstates participating in the evolution of the initial state determine whether self-averaging is enforced or not. 

In the particular case of initial coherent Gibbs states,
\begin{equation}
\overline{ R_{S_P^{\beta}}^{\kappa \neq 0} } = \frac{ \left\langle \dfrac{Z(2 \beta)^2 }{Z( \beta)^4} \right\rangle - \left\langle \dfrac{Z(2 \beta) }{Z( \beta)^2}  \right\rangle^2  }{ \left\langle \dfrac{Z(2 \beta) }{Z( \beta)^2}  \right\rangle^2 },
\label{Eqn:Relative_variance_of_SP_Gibbs}  
\end{equation}
which means that for infinite temperature,
\begin{equation}
\overline{ R_{S_P^{\beta=0}}^{\kappa \neq 0} } \rightarrow 0,
\label{Eqn:Relative_variance_of_infiniteTGibbs}
\end{equation}
and self-averaging is guaranteed for any system. This is not necessarily the case for coherent Gibbs states with finite temperature and for initial states in quench dynamics, as shown in the next sections.

Notice that for the isolated system, $\kappa=0$, the result for $\overline{R_{S_P} }$ changes significantly, because in Eq.~(\ref{Eq:SPsquare}) the term for $n=q$ and $m=p$ with $n\neq m$ and $p \neq q$ is also nonzero, which gives
\begin{equation}
\overline{ R_{S_P}^{\kappa=0} } = \frac{ \left\langle \sum_{n\neq m}|c_{n}^{(0)}|^{4}|c_{m}^{(0)}|^{4} \right\rangle }{  \left\langle \sum_{n}|c_{n}^{(0)}|^{4} \right\rangle^2  } .
\label{Eqn:RVisolated}  
\end{equation}
This equation leads to results that are entirely different from those of $\kappa \neq 0$ in Eq.~(\ref{Eqn: Relative variance of W --- long times}). Take, for example, the case of $\beta=0$ and $|c_n^{(0)}|^2 = 1/N$. We have that 
\begin{equation}
\overline{ R_{S_P^{\beta =0}}^{\kappa=0}}= \frac{ 1/N^2 - 1/N^3}{  1/N^2  } \approx 1,
\label{Eqn:RVisolated_beta0}  
\end{equation}
which is in stark contrast with the zero relative variance obtained in Eq.~(\ref{Eqn:Relative_variance_of_infiniteTGibbs}).


\begin{figure*}
\centering
\includegraphics[scale=0.55]{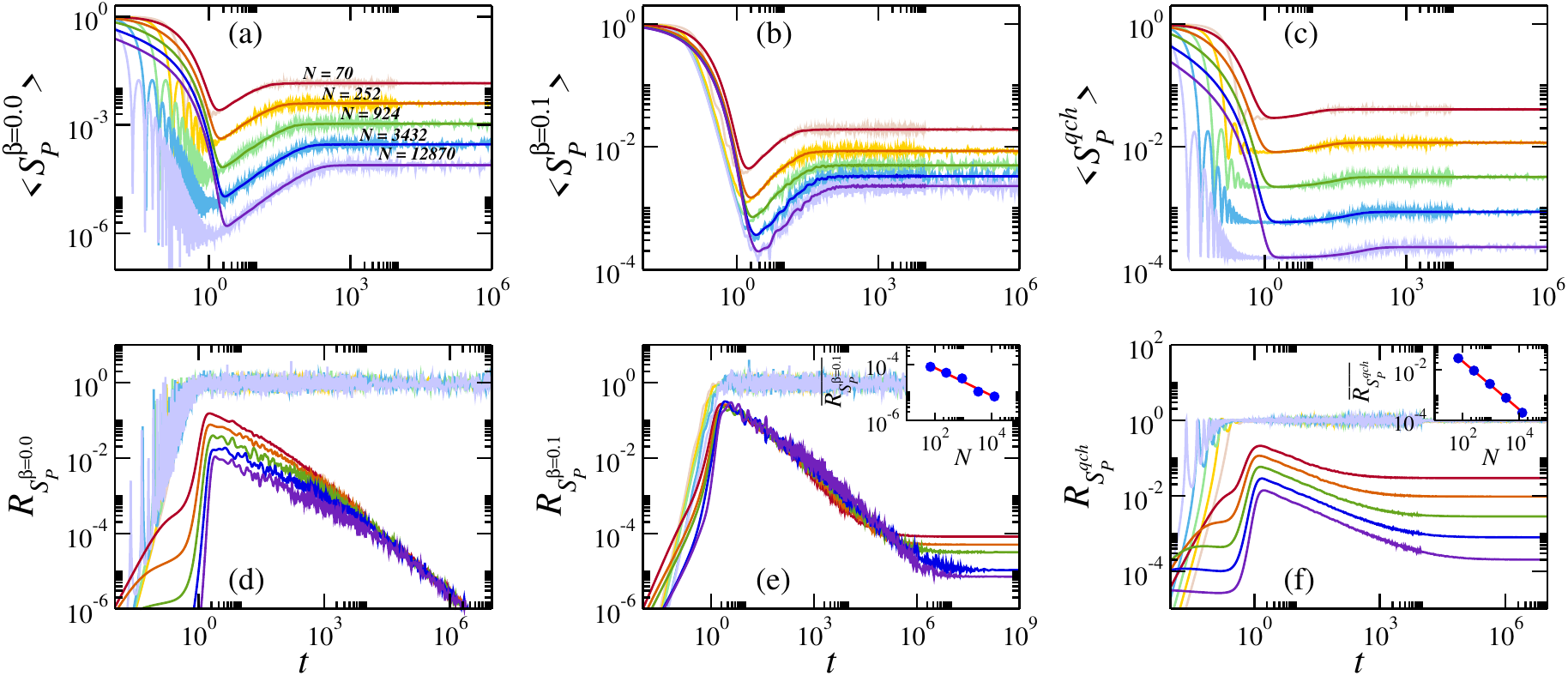}
\caption{Entire evolution of the average survival probabilities: (a) $ \langle S_{P}^{\beta=0.0} (t) \rangle$, (b) $\langle S_{P}^{\beta=0.1} (t) \rangle$, and (c) $ \langle S_{P}^{qch} (t) \rangle$, and their corresponding relative variances: (d) $R_{S_{P}^{\beta=0.0}} (t)$, (e) $R_{S_{P}^{\beta=0.1}} (t)$, and (f) $R_{S_{P}^{qch}} (t)$ for GOE random matrices.  Light colors correspond to the isolated model ($\kappa = 0$) and dark colors correspond to the open case ($\kappa = 0.05$). The inset of Fig.~\ref{Fig: Fig02. GOE. All}(e) [Fig.~\ref{Fig: Fig02. GOE. All}(f)]
shows the scaling analysis for $\overline{ R^{\kappa \neq 0}_{S_{P}^{\beta=0.1}}} $ [$\overline{R^{\kappa \neq 0}_{S_{P}^{qch}} }$] and the red line indicates the fitting $\overline{ R^{\kappa \neq 0}_{S_{P}^{\beta=0.1}} } \approx N^{-0.5}$ [$\overline{ R^{\kappa \neq 0}_{S_{P}^{qch}} }\approx 1.83\,N^{-0.96} $, in agreement with the analytical expression in Eq.~(\ref{Eq:analyticalRqch})].
}
\label{Fig: Fig02. GOE. All}
\end{figure*}

\section{Random Matrices}
\label{Sec: Models}

In this section, we analyze the evolution of the survival probability under full random matrices of the Gaussian orthogonal ensemble (GOE), which is a generic example of a fully chaotic system. We also consider PBRM and the RPE of random matrices, both of which allow us to explore what happens in the delocalized and localized regimes.

\subsection{Gaussian orthogonal ensemble}
\label{SubSec: GOE model}

Full random matrices are $N$-dimensional matrices filled with random numbers conditioned by general symmetry constraints. These matrices have been extensively used to reproduce the statistical properties of the spectra of complex quantum systems. The model is not physical because it implies the simultaneous interaction of all particles, but it allows for the identification of universal properties and the derivation of analytical results. 

In GOE, the random matrices are real and symmetric~\cite{MehtaBook}
and can be generated by adding a matrix $M$ filled with random numbers from a Gaussian distribution with mean $0$ and variance $1$  to its transpose,  $H=(M+M^{T})/2$. This means that the elements of $H$ have mean $\braket{H_{ij}}=0$ and variance
\begin{equation}
\braket{H_{ij}^{2}} = \left\lbrace
\begin{matrix}
1 ,& i=j,\\
1/2 ,& i\neq j.
\end{matrix}
\right.
 \label{Eqn: GOE model}
\end{equation}
To portray the case of quench dynamics, we assume that the initial Hamiltonian $H_0$ that defines the initial state is given by the diagonal part of the full random matrix.  The initial states are chosen close to the middle of the spectrum having energy $E^{(0)} =\langle \Psi(0)|H| \Psi(0) \rangle \sim 0$. Since the eigenstates of GOE matrices are random vectors, the components of the initial states are Gaussian random numbers with the constraint of normalization, so $ \langle |c_n^{(0)}|^2 \rangle \sim 1/N$.

In Fig.~\ref{Fig: Fig02. GOE. All}, we show the entire evolution under GOE matrices of $\langle S_{P}^{\beta=0} (t) \rangle$ [Fig.~\ref{Fig: Fig02. GOE. All}(a)] and its relative variance [Fig.~\ref{Fig: Fig02. GOE. All}(d)], of $\langle S_{P}^{\beta \neq0} (t) \rangle$ [Fig.~\ref{Fig: Fig02. GOE. All}(b)] and its relative variance [Fig.~\ref{Fig: Fig02. GOE. All}(e)], and of $\langle S_{P}^{qch}(t) \rangle$ [ Fig.~\ref{Fig: Fig02. GOE. All}(c)] and its relative variance [Fig.~\ref{Fig: Fig02. GOE. All}(f)]. 
Light colors are used for the isolated case ($\kappa=0$), and dark colors give the results for the open system ($\kappa=0.05$).
In addition to reducing the fluctuations throughout the dynamics, one sees that energy dephasing slows down the initial decay of the average survival probability in Fig.~\ref{Fig: Fig02. GOE. All}(a) and Fig.~\ref{Fig: Fig02. GOE. All}(c), suppressing the oscillations that are associated with the bounds of the energy distribution of the initial state~\cite{Tavora2016,Tavora2017}. We also observe that the saturation of the relative variance takes much longer to happen in Figs.~\ref{Fig: Fig02. GOE. All}(e)-(f) than the saturation of $\langle S_P (t) \rangle$ in Figs.~\ref{Fig: Fig02. GOE. All}(a)-(c).

For very short times, $R_{S_{P}^{\beta=0}} (t \rightarrow 0) $, $R_{S_{P}^{\beta \neq 0}} (t \rightarrow 0)$, and $R_{S_{P}^{qch}} (t \rightarrow 0)$ are very small, and according to Eq.~(\ref{Eqn:RVshortTime}), there is no self-averaging for isolated or open dynamics in GOE matrices, since $\sigma_{d_2}^2$ grows as $N$ increases. We verified this numerically, but it is not seen in the timescales of Figs.~\ref{Fig: Fig02. GOE. All}(d)-(f).

For long times, beyond the correlation hole of $\langle S_P (t) \rangle$, the relative variances for the isolated GOE model in Figs.~\ref{Fig: Fig02. GOE. All}(d)-(f) go to 1 for any large $N$. This is in agreement with Eq.~(\ref{Eqn:RVisolated_beta0}) and indicates the lack of self-averaging. This picture is reversed for the open system in all three cases: $\overline{ R_{S_{P}^{\beta=0}}^{\kappa \neq 0}  } $ in Fig.~\ref{Fig: Fig02. GOE. All}(d) goes to zero, as justified with the derivation of Eq.~(\ref{Eqn:Relative_variance_of_infiniteTGibbs}), and both  $\overline{ R_{S_{P}^{\beta \neq 0}}^{\kappa \neq 0}  }$ and $\overline{ R_{S_{P}^{qch}}^{\kappa \neq 0}  }$ are $ \propto 1/N$, indicating ``super'' self-averaging~\cite{Schiulaz2020}, as shown in the insets of Figs.~\ref{Fig: Fig02. GOE. All}(e)-(f). 

The behavior $\overline{ R_{S_{P}^{qch}}^{\kappa \neq 0}  } \propto 1/N$ can be explained using Eq.~(\ref{Eqn: Relative variance of W --- long times}) and the fact that $c_n^{(0)}$ are Gaussian random numbers satisfying $\sum_n |c_n^{(0)}|^2=1$. Taking into account that different eigenstates of GOE matrices are statistically independent, we have that
$\braket{|c_{n}^{(0)}|^{4}|c_{m}^{(0)}|^{4}}=\braket{|c_{n}^{(0)}|^{4}}\braket{|c_{m}^{(0)}|^{4}}$ for $n\neq m$, which leads to
\begin{align}
\nonumber \sigma_{\text{IPR}_0}^{2}&=\braket{\sum_{n,m=1}^{N}|c_{n}^{(0)}|^{4}|c_{m}^{(0)}|^{4}}-\braket{\sum_{n=1}^{N}|c_{n}^{(0)}|^{4}}\braket{\sum_{m=1}^{N}|c_{m}^{(0)}|^{4}} \\
\nonumber &=\sum_{n,m=1}^{N}\left[\braket{|c_{n}^{(0)}|^{4}|c_{m}^{(0)}|^{4}}-\braket{|c_{n}^{(0)}|^{4}}\braket{|c_{m}^{(0)}|^{4}}\right] 
\\
 &=\sum_{n=1}^{N}\left[\braket{|c_{n}^{(0)}|^{8}}-\braket{|c_{n}^{(0)}|^{4}}^{2}\right] \sim \mathcal{O}(N^{-3}).
\label{Eqn: Sp variance. Open system. GOE}  
\end{align}
At the same time, $\langle \text{IPR}_0 \rangle \sim 3/N$, so 
\begin{equation}
    \overline{ R_{S_{P}^{qch}}^{\kappa \neq 0} } \propto 1/N ,
    \label{Eq:analyticalRqch}
\end{equation}
as confirmed with the inset of Fig.~\ref{Fig: Fig02. GOE. All}(f).

The contrasting results for $\overline{R_{S_P}}$ for closed and open systems can also be understood from the analysis of the distribution of the values of $S_P(t)$ at times after the saturation of the dynamics. For the isolated system, the distribution is exponential, so the width coincides with the mean~\cite{Torres2020PRE} and $\overline{R_{S_P}^{\kappa =0} } =1$. By opening the system, the distribution approaches a delta function for $S_{P}^{\beta=0} (t \rightarrow \infty)$, Gaussian for $S_{P}^{\beta=0.1} (t \rightarrow \infty)$, and Fr\'{e}chet-like for $S_{P}^{qch} (t \rightarrow \infty)$, so the width can decrease with respect to the mean, which allows for the decay of the relative variance as $N$ increases. 
 
The behavior of the relative variance in the region of the correlation hole is also worth noting. At the timescale where the ramp starts, $R_{S_P}(t)$ rises to its largest values for the isolated and open models because $\langle S_P(t) \rangle$ reaches its minimum value. Beyond this point, $R_{S_P}^{\kappa \neq 0}(t)$ decreases monotonically (up to saturation), while $R_{S_P}^{\kappa = 0}(t)$ saturates to 1. 

In a nutshell, apart from very short times, self-averaging holds throughout the evolution of the open systems in Fig.~\ref{Fig: Fig02. GOE. All}(d) and Fig.~\ref{Fig: Fig02. GOE. All}(f).  In Fig.~\ref{Fig: Fig02. GOE. All}(e), where the initial state is too close to the edge of the spectrum, the decay of $R_{S_P^{\beta=0.1}}^{\kappa \neq 0} (t)$ with $N$ is evident only after saturation. We deepen the discussion on how $R_{S_P}^{\kappa \neq 0} (t)$ can depend on the choice of the initial state in the next two sections. 

\begin{figure*}
\centering
\includegraphics[scale=0.6]{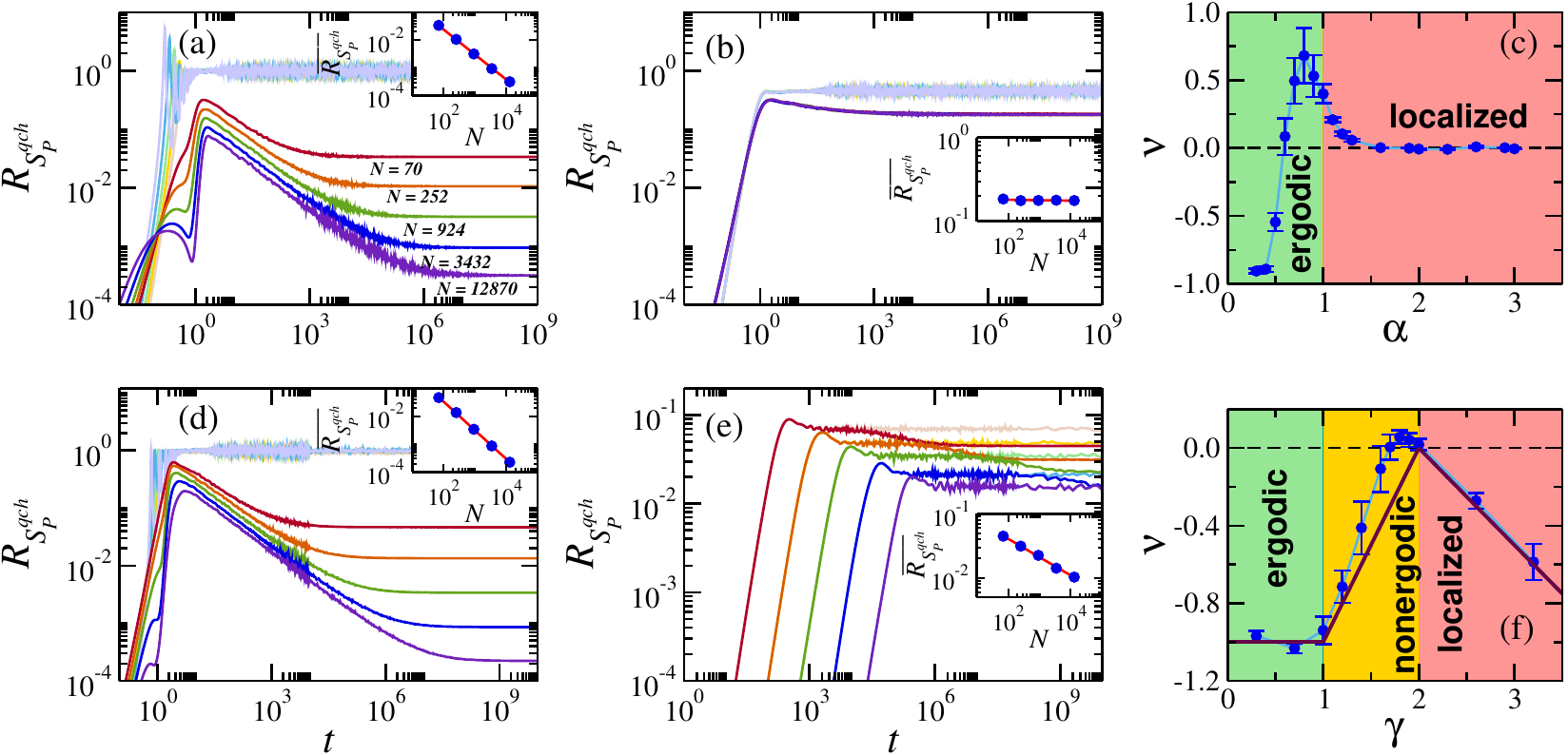}
\caption{Entire evolution of $R_{S_{P}^{qch}}(t)$ under quench dynamics for (a,b) the PBRM model and (d,e) the RPE. Panels (a,d) are for the delocalized regime ($\alpha=0.3$, $\gamma=0.7$) and panels (b,e) are for the localized regime ($\alpha=3.0$, $\gamma=2.6$). Light colors are for the isolated model, and dark colors are for the open model.  The scaling analysis in the inset of panel (a) gives the fitting 
$\overline{R^{\kappa \neq 0}_{S_P^{qch}}}=1.52\,N^{-0.90}$, in panel (d)  $\overline{R^{\kappa \neq 0}_{S_P^{qch}}}=3.83\,N^{-1.0}$, and 
in panel (e) 
$\overline{R^{\kappa \neq 0}_{S_P^{qch}}}=0.15\,N^{-0.29}$. Panel (c) for the PBRM model [panel (f) for the RPE] gives the exponent $\nu$ obtained from the scaling analysis of $\overline{R^{\kappa \neq 0}_{S_{P}^{qch}}} \propto N^{\nu}$ as a function of the control parameter $\alpha$ [$\gamma$]. The thin blue line in panels (c) and (f) guides the eye, the thick maroon line in panel (f) corresponds to Eq.~(\ref{Eqn: RPE RV scaling}).
}
\label{Fig: Fig03. PBRM and RPE models}
\end{figure*}

\subsection{Power-law banded random matrices}
\label{SubSubSec: PBRM}

In the PBRM ensemble~\cite{PhysRevE.54.3221,PhysRevB.61.R11859} the elements of the matrices are Gaussian random numbers with $\braket{H_{ij}}=0$ and variance 
\begin{equation}
\left\langle  H_{ij}^2\right\rangle=
\begin{cases}
    1,& i=j,\\
    \left(1+ \left| i-j \right|^{2 \alpha} \right)^{-1},& i\neq j,
\end{cases}
\label{Eqn: PRMB elements}
\end{equation}
where $\alpha\in\left(0,\infty\right)$ is a control parameter. The model shows a phase transition determined by the value of $\alpha$. For $\alpha<1$, the system is in the chaotic (delocalized) regime, while for $\alpha>1$, the system is in the localized regime. The ensemble has two limiting cases: if $\alpha\rightarrow 0$, we obtain the GOE model, and if $\alpha \rightarrow\infty$, we have a tridiagonal matrix.

In Figs.~\ref{Fig: Fig03. PBRM and RPE models}(a)-(b), we show results for the relative variance of the survival probability of the PBRM model in the delocalized phase [$\alpha=0.3$ in Fig.~\ref{Fig: Fig03. PBRM and RPE models}(a)] and in the localized phase [$\alpha=3$ in Fig.~\ref{Fig: Fig03. PBRM and RPE models}(b)]. We consider the case of quench dynamics, where the initial Hamiltonian $H_0$ is the diagonal part of the PBRM. The initial states have energy $E^{(0)}$ in the middle of the spectrum. Light colors are for the isolated model and dark colors are for the open system.

In the delocalized phase of the open PBRM model, there is lack of self-averaging only at very short times, in agreement with Eq.~(\ref{Eqn:RVshortTime}), but soon the curves cross, ensuring self-averaging throughout the dynamics. The scaling analysis in the inset of Fig.~\ref{Fig: Fig03. PBRM and RPE models}(a) indicates that $\overline{R^{\kappa \neq 0}_{S_P^{qch}}} \propto N^{-0.9}$, similar to what was found for the GOE model in the inset of Fig.~\ref{Fig: Fig02. GOE. All}(f).

The results for the localized phase in Fig.~\ref{Fig: Fig03. PBRM and RPE models}(b) are very different from the delocalized phase. Even though the relative fluctuations are reduced by opening the system,  $R^{\kappa \neq 0}_{S_P^{qch}}(t) < R^{\kappa = 0}_{S_P^{qch}}(t)$, the relative variance for $\kappa \neq 0$ does not decrease as $N$ increases. The inset in Fig.~\ref{Fig: Fig03. PBRM and RPE models}(b) makes it clear that $\overline{R^{\kappa \neq 0}_{S_P^{qch}}}$ as a function of $N$ is approximately constant. This contrasts with the case of a coherent Gibbs state with $\beta=0.1$, for which $\overline{R^{\kappa \neq 0}_{S_P^{\beta=0.1}}}$ in the localized phase does decrease with $N$ (not shown). These different behaviors indicate the important role of the initial state and the varying degrees of fluctuations at different parts of the spectrum.

In Fig.~\ref{Fig: Fig03. PBRM and RPE models}(c), we analyze the dependence of the self-averaging properties of $S_P^{qch}(t)$ on the control parameter $\alpha$. We focus on initial states in the middle of the spectrum and the relative variance after saturation. We perform scaling analysis of 
\begin{equation}
  \overline{R^{\kappa \neq 0}_{S_P^{qch}}} \propto N^{\nu}  
  \label{Eq:scalingR}
\end{equation}
and show $\nu$ as a function of $\alpha$. If $\nu \geq0 $, then $S_P^{qch}(t \rightarrow \infty)$ is non-self-averaging, and it is self-averaging otherwise.  Figure~\ref{Fig: Fig03. PBRM and RPE models}(c) confirms that self-averaging holds for initial states with $E^{(0)} \sim 0$ only in the delocalized regime. We observe the following behavior:

(i) The exponent $\nu<0$ for $\alpha \lesssim 1/2$, where the PBRM ensemble is similar to the GOE \cite{PhysRevE.54.3221}.

(ii) As $\alpha$ approaches the critical point $\alpha_{c}=1$, we see that $\nu$ grows significantly, indicating a strong lack of self-averaging. This is consistent with expectations that, at a critical point, the structures of the states vary considerably. The exponent $\nu$ attains its maximum close to the delocalization-localization transition point. 

(iii) The value of $\nu$ then decreases in the localized phase. For $1 < \alpha \lesssim 3/2$, where super-diffusive dynamics is observed at short time, $\nu$ is still positive. The scaling exponent $\nu\to 0$ for $\alpha >3/2$, where the lack of self-averaging may be attributed to the power-law tails of the localized eigenstates.

\subsection{Rosenzweig-Porter ensemble}
\label{SubSubSec: RPE}

The RPE was first introduced to explain the level statistics of heavy atoms~\cite{Rosenzweig1960}. An $N\times N$ matrix from the real symmetric RPE has random elements from a Gaussian distribution with mean $0$ and variance 
\begin{equation}
\langle H_{ij}^2\rangle=
\begin{cases}
    1, & i=j,\\
    \dfrac{1}{2N^\gamma}, & i\neq j ,
\end{cases} 
\label{Eqn: RPE elements}
\end{equation}
where the system parameter $\gamma\in\mathbb{R}$. The RPE is essentially a Poisson ensemble perturbed by the GOE, hence, a deformed ensemble~\cite{Carvalho2007, Das2022a, Das2022b} that mimics how the symmetries of an integrable system represented by the Poisson ensemble are gradually broken as $\gamma$ decreases from infinity. The RPE hosts three distinct phases: an ergodic (chaotic) phase ($\gamma<1$), a nonergodic extended phase having fractal eigenstates ($1<\gamma<2$), and a localized phase ($\gamma>2$) \cite{Kravtsov2015, Monthus2017, Pino2019, Soosten2019}. These phase transitions have been explored experimentally in microwave resonators~\cite{Zhang2023}. Non-trivial fractal phases similar to that in the RPE have been observed in other random matrix models~\cite{Nosov2019, Cizeau1994, Kutlin2021, Tang2022, Motamarri2021, Das2022c, Das2023a, Das2024a}, hierarchical graphs~\cite{Tikhonov2016}, and many-body disordered systems~\cite{Luitz2015, Luitz2020, Pino2017}. The fractal states at the delocalization-localization transition point of the RPE have the same statistical properties as those in hierarchical lattices, such as the Bethe lattice or random regular graphs~\cite{Kravtsov2015}. Since the Fock space of generic isolated many-body quantum systems has a hierarchical structure~\cite{Altshuler1997}, the RPE has gained a lot of attention in recent times~\cite{Khaymovich2020, Khaymovich2021, Tomasi2019, Bogomolny2018, Facoetti2016, Venturelli2023, Das2019, Das2023b}.

In Figs.~\ref{Fig: Fig03. PBRM and RPE models}(d)-(e), we show results for the relative variance of the survival probability of the RPE in the ergodic phase [$\gamma=0.7$ in Fig.~\ref{Fig: Fig03. PBRM and RPE models}(d)] and in the localized phase [$\gamma=2.6$ in Fig.~\ref{Fig: Fig03. PBRM and RPE models}(e)]. We consider quench dynamics, where the initial state comes from the diagonal part of the RPE and has energy $E^{(0)}$ in the middle of the spectrum. Light colors are for the isolated model, and dark colors are for the open system.

The results for the ergodic regime of RPE in Fig.~\ref{Fig: Fig03. PBRM and RPE models}(d) are analogous to those for the PBRM model in Fig.~\ref{Fig: Fig03. PBRM and RPE models}(a), namely, by opening the chaotic system, we induce self-averaging and at long times $\overline{R^{\kappa \neq 0}_{S_P^{qch}}} \propto 1/N$, as in the GOE case. Less expected are the results in the localized phase in  Fig.~\ref{Fig: Fig03. PBRM and RPE models}(e), where self-averaging holds for all timescales shown in the figure, even for the isolated model. Furthermore, the effects of the energy dephasing in $R^{\kappa \neq 0}_{S_P^{qch}}(t)$ develop only at very long times that increase as $N$ increases. This suggests that the structures of the eigenstates at different energies are very similar in the localized phase of the RPE.

\begin{figure*}
\centering
\includegraphics[scale=0.6]{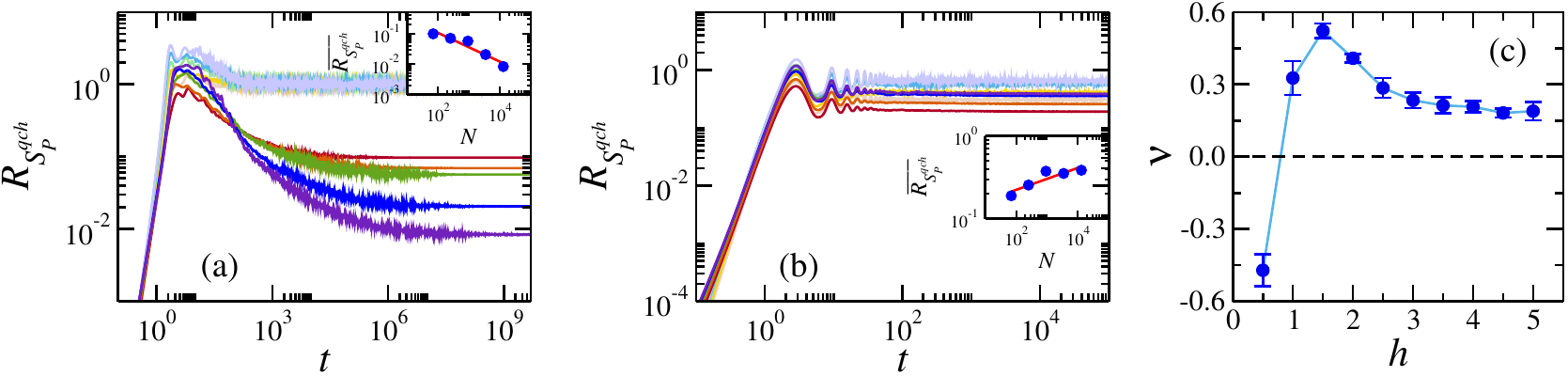}
\caption{Entire evolution of $R_{S_{P}^{qch}}(t)$ under the disordered XXX model in (a) the chaotic regime ($h=0.5$) and (b) the localized regime ($h=5.0$).  Light colors are for the isolated model and dark colors are for the open model. From top to bottom at long times in (a), the dark lines correspond to the following system sizes:
 $L=8 \rightarrow N=70$, $L=10 \rightarrow N=252$, $L=12 \rightarrow N=924$, $L=14 \rightarrow N=3432$, and $L=16 \rightarrow N=12870$.
The scaling analysis in the inset of panel (a) gives the fitting 
$\overline{R^{\kappa \neq 0}_{S_P^{qch}}}=0.92\,N^{-0.47}$,
and in panel (b)  $\overline{R^{\kappa \neq 0}_{S_P^{qch}}}=0.12\,N^{0.13}$. 
Panel (c) gives the power $\nu$, obtained from the scaling analysis of $\overline{R^{\kappa \neq 0}_{S_{P}^{qch}}} \propto N^{\nu}$, as a function of the disorder strength $h$.
}
\label{Fig: Fig05. Disordered XXX, E middle}
\end{figure*}

The results for the scaling analysis in Eq.~(\ref{Eq:scalingR}) for various values of $\gamma$ are given in Fig.~\ref{Fig: Fig03. PBRM and RPE models}(f) and indicate that  
\begin{equation}
\nu \approx
\begin{cases}
    -1, & \gamma\leq 1,\\
    \gamma-2, & 1 < \gamma < 2,\\
    1 - \gamma/2, & \gamma \geq 2.
\end{cases} 
\label{Eqn: RPE RV scaling}
\end{equation}
We thus have the following picture:

(i) In the ergodic regime ($\gamma < 1$), the bulk eigenstates are like random vectors and the asymptotic relative variance scales as $1/N$, similar to GOE, so $\nu \approx -1$.

(ii) In the nonergodic regime of RPE ($1< \gamma < 2$), the leading portion of the spectrum consists of eigenstates with $\langle \text{IPR} \rangle \propto N^{-D}$, where $D$ is the fractal exponent given by $D\approx 2-\gamma$ \cite{Kravtsov2015}. Similarly, the asymptotic survival probability of initial states in the middle of the spectrum scales as $N^{-D}$ \cite{Tomasi2019}. We find numerically that the asymptotic relative variance also scales approximately as $N^{-D}$, therefore $\nu \approx \gamma -2$.

(iii) At $\gamma = 2$, where RPE exhibits a second-order phase transition from the delocalized to the localized phase, the correlation length diverges in a power-law manner and the  density of bulk localization lengths becomes scale invariant. Hence, the variance of IPR is $\mathcal{O}(1)$ while the IPR itself is also $\mathcal{O}(1)$, leading to $\nu\approx 0$.
Similar to what is seen for the PBRM in Fig.~\ref{Fig: Fig03. PBRM and RPE models}(c), at the delocalization-localization transition point, energy dephasing is unable to enforce self-averaging.

(iv) In the localized regime ($\gamma>2$), the fractal dimension $D=0$ \cite{Kravtsov2015}, so
the mean asymptotic survival probability is ${\cal O}(1)$,  while the fluctuation in the IPR$_0$ is dictated by the small components at the tail of the states, leading to $\nu \approx 1-\gamma/2$.
Consequently, $\nu<0$, indicating the presence of self-averaging.

In short, by opening the system, the survival probability for initial states in the middle of the spectrum becomes self-averaging for the PBRM model and the RPE in the chaotic regime. The method fails at the localization transition critical point for both systems and in the localized phase for the PBRM model.


\section{Physical system: spin-1/2 Heisenberg model}
\label{Sec: Physical models}

The analysis of self-averaging done above for different kinds of random matrices sets the scene for the study of self-averaging in physical models. In particular, the previous discussions anticipate differences in ergodic and nonergodic phases and a possible dependence on the energy of the initial states. 

We consider the one-dimensional isotropic spin-1/2 Heisenberg with onsite disorder (also known as disordered XXX model), which has been extensively investigated in the context of many-body localization both theoretically and experimentally~\cite{SantosEscobar2004,Dukesz2009,Pal2010,Schreiber2015,Luitz2017,Sierant2024ARXIV}. The Hamiltonian is
\begin{equation}
 H=J\sum_{k=1}^{L}h_{k}S_{k}^{z} + J\sum_{k=1}^{L-1}(S_{k}^{x}S_{k+1}^{x}+S_{k}^{y}S_{k+1}^{y}+S_{k}^{z}S_{k+1}^{z}),
 \label{Eqn: XXXh model}
\end{equation}
where $\hbar=1$, $J=1$ is the coupling strength, $S_{k}^{x,y,z}$ are spin operators on site $k$, $L$ is the size of the chain, and open boundary conditions are used. The Zeeman splittings $h_{k}$ are random numbers uniformly distributed in $[-h,h]$ and $h$ is the disorder strength. The total magnetization in the $z$-direction, ${\cal S}^z = \sum_{k=1}^L S_k^z$, is conserved, so we take the largest subspace, where ${\cal S}^z=0$ and the dimension is $N=L!/(L/2)!^2$. The model is integrable when $h=0$ and becomes chaotic for $0<h\lesssim 1$. Whether the system in the thermodynamic limit can reach a many-body localized phase when $h$ is above some critical value larger than the coupling strength has been debated. Despite the controversy, our numerical studies are performed with finite systems, so we refer to a ``localized phase'' for large $h$.

We start the analysis of self-averaging in Sec.~\ref{SubSec: Middle of the spectrum} with initial states corresponding to eigenstates of $H_0=J\sum_{k=1}^{L}h_{k}S_{k}^{z} + J\sum_{k=1}^{L} S_{k}^{z}S_{k+1}^{z}$ that have energy $E^{(0)}$ in the middle of the spectrum. Various values of $h$ are considered. In Sec.~\ref{SubSec: Edge of the spectrum}, we investigate how the choice of initial state affects the results for the chaotic and localized phases.

\subsection{Middle of the spectrum}
\label{SubSec: Middle of the spectrum}

In Figs.~\ref{Fig: Fig05. Disordered XXX, E middle}(a)-(b), we show the relative variance of the survival probability for the disordered spin-1/2 model in the chaotic regime [Fig.~\ref{Fig: Fig05. Disordered XXX, E middle}(a)] and in the localized phase [Fig.~\ref{Fig: Fig05. Disordered XXX, E middle}(b)], where the initial state has energy $E^{(0)}$ in the middle of the spectrum. Light colors are for the isolated model, and dark colors are for the open system. 

Deep in the chaotic regime ($h\lesssim 1$), the results are comparable to those for the random matrices in Fig.~\ref{Fig: Fig02. GOE. All}(f) and in Fig.~\ref{Fig: Fig03. PBRM and RPE models}(a,d). The results follow the generic picture, that is, in the chaotic phase and for initial states projected into chaotic energy eigenstates, the survival probability in the timescales of the correlation hole and beyond becomes self-averaging if we open the system to a dephasing environment. Notice, however, that $\nu$ is negative in  the inset of  Fig.~\ref{Fig: Fig05. Disordered XXX, E middle}(a), but does not reach  -1 as in GOE. This is because even the highly excited eigenstates in the middle of the spectrum are not ergodic in the sense of Haar measure~\cite{Luitz2017}.

The onset of self-averaging could facilitate the experimental detection of the correlation hole~\cite{Das2024}, since in the presence of a dephasing environment, one could reduce the number of initial states and disorder realizations for the experimental studies. But the dephasing strength should not be too large to avoid suppressing  the correlation hole \cite{Xu2019}.

In the localized phase, the fluctuations are large, and self-averaging is not achieved through the environment. The scenario is even worse than in the PBRM model in Fig.~\ref{Fig: Fig03. PBRM and RPE models}(b), since for the spin model in Fig.~\ref{Fig: Fig05. Disordered XXX, E middle}(b), $R^{\kappa \neq 0}_{S_P^{qch}}(t)$ increases with $N$. This behavior is also shown for the asymptotic value of the relative variance  in the inset of Fig.~\ref{Fig: Fig05. Disordered XXX, E middle}(b). Based on the available system sizes, we cannot say whether $\overline{R^{\kappa \neq 0}_{S_P^{qch}}}$ will eventually converge to $\overline{R^{\kappa = 0}_{S_P^{qch}}}$, but the results point to the difficulties associated with the numerical analysis of many-body localization. The problem is not only the exponentially large Hilbert space in $L$ but also the large fluctuations that one has to deal with~\cite{Solorzano2021}.

In Fig.~\ref{Fig: Fig05. Disordered XXX, E middle}(c), we plot $\nu$ obtained with Eq.~(\ref{Eq:scalingR}) as a function of the disorder strength $h$. There is a parallel with the analysis for the PBRM model in Fig.~\ref{Fig: Fig03. PBRM and RPE models}(c), in the sense that both models do not show self-averaging away from the chaotic regime, but the results are worse for the spin model. In this case, $\nu>0$ for any $h >1$, while for the PBRM model, $\nu>0$ only in the vicinity of the critical point. 

The comparison between Fig.~\ref{Fig: Fig03. PBRM and RPE models}(c), Fig.~\ref{Fig: Fig03. PBRM and RPE models}(f), and
Fig.~\ref{Fig: Fig05. Disordered XXX, E middle}(c) fuels speculation of the role of $\nu$ in the study of many-body localization. For example, similar to the PBRM and RPE, the exponent $\nu$ for the disordered spin model also reaches a maximum value at a point that could be associated with a transition. 
However, given the limited system sizes available for numerical studies, it is difficult to further elaborate on this topic.


\subsection{Edge of the spectrum}
\label{SubSec: Edge of the spectrum}

\begin{figure}[h]
\centering
\includegraphics[scale=0.55]{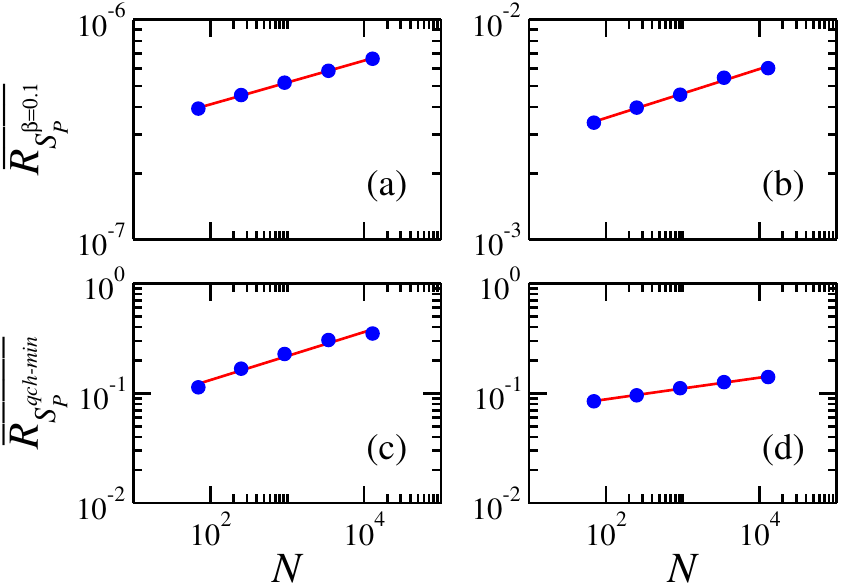}
\caption{Infinite-time average of the relative variance, (a,b) $\overline{R^{\kappa \neq 0}_{S_{P}^{\beta=0.1}}}$ and (c,d) $\overline{R^{\kappa \neq 0}_{S_{P}^{qch-min}}}$, as a function of the Hilbert space size. The evolution is performed under the disordered XXX model in (a,c) the chaotic regime and (b,d) the localized phase. Contrary to Fig.~\ref{Fig: Fig05. Disordered XXX, E middle}, the initial state in the quench dynamics has the lowest energy. Fittings: 
(a) $\overline{R^{\kappa \neq 0}_{S_{P}^{\beta=0.1}}} = 2.5\times10^{-7} N^{0.10} $
(b) $\overline{R^{\kappa \neq 0}_{S_{P}^{\beta=0.1}}} = 2.3\times10^{-3} N^{0.10} $
(c) $\overline{R^{\kappa \neq 0}_{S_{P}^{qch-min}}} = 4.7\times10^{-2} N^{0.22}$, and
(d) $\overline{R^{\kappa \neq 0}_{S_{P}^{qch-min}}} = 5.7\times10^{-2} N^{0.09}$.
}
\label{Fig: Fig06. Disordered XXX model. Emin}
\end{figure}

Motivated by the discussion in Sec.~\ref{SubSubSec: PBRM} about the different behavior with $N$ of $\overline{R^{\kappa \neq 0}_{S_P^{qch}}}$ (initial state energy in the middle of the spectrum) and $\overline{R^{\kappa \neq 0}_{S_P^{\beta=0.1}}}$ (initial state energy at the low edge of the spectrum) for the PBRM model in the localized phase, we now investigate how the results presented in Fig.~\ref{Fig: Fig05. Disordered XXX, E middle} for the spin model are affected by the choice of initial states. For this study, we consider only initial states with very low energies, so we refer to the asymptotic relative variance obtained for quench dynamics as $\overline{R^{\kappa \neq 0}_{S_P^{qch-min}}}$.

In Fig.~\ref{Fig: Fig06. Disordered XXX model. Emin}, we present a scaling analysis of the infinite-time average of the relative variance for the initial coherent Gibbs state with $\beta=0.1$ for the spin model in the chaotic [Fig.~\ref{Fig: Fig06. Disordered XXX model. Emin}(a)] and localized phase [Fig.~\ref{Fig: Fig06. Disordered XXX model. Emin}(b)] and for the initial state quenched in the lowest-energy part of the spectrum for the spin model in the chaotic [Fig.~\ref{Fig: Fig06. Disordered XXX model. Emin}(c)] and localized phase [Fig.~\ref{Fig: Fig06. Disordered XXX model. Emin}(d)]. All panels manifest a lack of self-averaging. The exponent $\nu$ is small but positive in all cases. The worst scenario happens in the chaotic regime for the quench dynamics at the edge of the spectrum [Fig.~\ref{Fig: Fig06. Disordered XXX model. Emin}(c)]. Spectral correlations as in random matrix theory and chaotic eigenstates are only present in the bulk of the spectrum of chaotic many-body quantum systems, while the edges exhibit states with highly fluctuating structures~\cite{Santos2010PRE,Torres2013}. This is related with the density of states of many-body systems with two-body couplings, which is Gaussian~\cite{French1970}, thus implying that the states closer to the edges are more localized.


\section{Conclusions}
\label{Sec: Conclusions}

Opening many-body quantum systems to a dephasing environment reduces dynamical fluctuations. We explored this fact to analyze the conditions under which the survival probability (equivalently, the spectral form factor) in physical systems becomes self-averaging. 

We started the study with different kinds of random matrices to identify general patterns. We showed that in full random matrices from the Gaussian orthogonal ensemble, the environment ensures self-averaging for any initial state, while in power-law banded random matrices and Rosenzweig-Porter random matrices, where a transition to a localized phase exists, self-averaging cannot be achieved at the critical point. Furthermore, the survival probability remains non-self-averaging for power-law banded random matrices in the localized phase due to the power-law tails of the localized eigenstates.

In agreement with the results above, the survival probability becomes self-averaging for an open disordered spin-1/2 model in the chaotic regime and for initial states in the middle of the spectrum. Any change in the regime or in the initial state can affect the outcome. Self-averaging does not hold for any initial state when the spin model ceases to be chaotic, and the same happens even in the chaotic regime if the initial state is away from the middle of the spectrum.

Being able to ensure self-averaging at long times is directly determined by the structures of the eigenstates in the spectrum region of the initial state energy. This is because the asymptotic relative variance of the survival probability in open systems, $\overline{R^{\kappa \neq 0}_{SP} }$, coincides with the relative variance of the inverse participation ratio of the initial state [cf. Eq.~(\ref{Eqn: Relative variance of W --- long times})].  Lack of self-averaging indicates that the structures of the states  vary significantly, as it happens at critical points. Thus, the behavior of $\overline{R^{\kappa \neq 0}_{SP} }$ works as a probe of the structure of the eigenstates in different parts of the spectrum and strong lack of self-averaging suggests the presence of critical points.  

The analysis in this work indicates that slightly opening the system to a dephasing environment should help with the experimental detection of many-body quantum chaos in complex systems out of equilibrium~\cite{Das2024}. This is because the environment ensures self-averaging in the timescales where the correlation hole emerges.

\begin{acknowledgments} 
	A.K.D. is supported by the Fulbright-Nehru Grant No.~2879/FNDR/2023-2024. D.A.Z.-H. and E.J.T.-H. are supported by CONAHCYT through project No. CF-2023-I-1748. E.J.T.-H. is  grateful for financial support from VIEP-BUAP, project No. 00427-2024. D.A.Z.-H. and E.J.T.-H. are grateful to LNS-BUAP for their supercomputing facility. 
 L.F.S. is supported by the NSF Grant No. DMR-1936006. This research was supported in part by the International Centre for Theoretical Sciences (ICTS) for participating in the program -  Periodically and quasi-periodically driven complex systems  (code: ICTS/pdcs2023/6).
 We thank Adolfo del Campo for several suggestions to the manuscript.
\end{acknowledgments}



\begin{thebibliography}{108}%
\makeatletter
\providecommand \@ifxundefined [1]{%
 \@ifx{#1\undefined}
}%
\providecommand \@ifnum [1]{%
 \ifnum #1\expandafter \@firstoftwo
 \else \expandafter \@secondoftwo
 \fi
}%
\providecommand \@ifx [1]{%
 \ifx #1\expandafter \@firstoftwo
 \else \expandafter \@secondoftwo
 \fi
}%
\providecommand \natexlab [1]{#1}%
\providecommand \enquote  [1]{``#1''}%
\providecommand \bibnamefont  [1]{#1}%
\providecommand \bibfnamefont [1]{#1}%
\providecommand \citenamefont [1]{#1}%
\providecommand \href@noop [0]{\@secondoftwo}%
\providecommand \href [0]{\begingroup \@sanitize@url \@href}%
\providecommand \@href[1]{\@@startlink{#1}\@@href}%
\providecommand \@@href[1]{\endgroup#1\@@endlink}%
\providecommand \@sanitize@url [0]{\catcode `\\12\catcode `\$12\catcode
  `\&12\catcode `\#12\catcode `\^12\catcode `\_12\catcode `\%12\relax}%
\providecommand \@@startlink[1]{}%
\providecommand \@@endlink[0]{}%
\providecommand \url  [0]{\begingroup\@sanitize@url \@url }%
\providecommand \@url [1]{\endgroup\@href {#1}{\urlprefix }}%
\providecommand \urlprefix  [0]{URL }%
\providecommand \Eprint [0]{\href }%
\providecommand \doibase [0]{https://doi.org/}%
\providecommand \selectlanguage [0]{\@gobble}%
\providecommand \bibinfo  [0]{\@secondoftwo}%
\providecommand \bibfield  [0]{\@secondoftwo}%
\providecommand \translation [1]{[#1]}%
\providecommand \BibitemOpen [0]{}%
\providecommand \bibitemStop [0]{}%
\providecommand \bibitemNoStop [0]{.\EOS\space}%
\providecommand \EOS [0]{\spacefactor3000\relax}%
\providecommand \BibitemShut  [1]{\csname bibitem#1\endcsname}%
\let\auto@bib@innerbib\@empty
\bibitem [{\citenamefont {Xu}\ \emph {et~al.}(2019)\citenamefont {Xu},
  \citenamefont {Garc\'{\i}a-Pintos}, \citenamefont {Chenu},\ and\
  \citenamefont {del Campo}}]{Xu2019}%
  \BibitemOpen
  \bibfield  {author} {\bibinfo {author} {\bibfnamefont {Z.}~\bibnamefont
  {Xu}}, \bibinfo {author} {\bibfnamefont {L.~P.}\ \bibnamefont
  {Garc\'{\i}a-Pintos}}, \bibinfo {author} {\bibfnamefont {A.}~\bibnamefont
  {Chenu}},\ and\ \bibinfo {author} {\bibfnamefont {A.}~\bibnamefont {del
  Campo}},\ }\bibfield  {title} {\bibinfo {title} {Extreme decoherence and
  quantum chaos},\ }\href {https://doi.org/10.1103/PhysRevLett.122.014103}
  {\bibfield  {journal} {\bibinfo  {journal} {Phys. Rev. Lett.}\ }\textbf
  {\bibinfo {volume} {122}},\ \bibinfo {pages} {014103} (\bibinfo {year}
  {2019})}\BibitemShut {NoStop}%
\bibitem [{\citenamefont {Cornelius}\ \emph {et~al.}(2022)\citenamefont
  {Cornelius}, \citenamefont {Xu}, \citenamefont {Saxena}, \citenamefont
  {Chenu},\ and\ \citenamefont {del Campo}}]{Cornelius2022}%
  \BibitemOpen
  \bibfield  {author} {\bibinfo {author} {\bibfnamefont {J.}~\bibnamefont
  {Cornelius}}, \bibinfo {author} {\bibfnamefont {Z.}~\bibnamefont {Xu}},
  \bibinfo {author} {\bibfnamefont {A.}~\bibnamefont {Saxena}}, \bibinfo
  {author} {\bibfnamefont {A.}~\bibnamefont {Chenu}},\ and\ \bibinfo {author}
  {\bibfnamefont {A.}~\bibnamefont {del Campo}},\ }\bibfield  {title} {\bibinfo
  {title} {Spectral filtering induced by non-{H}ermitian evolution with
  balanced gain and loss: Enhancing quantum chaos},\ }\href
  {https://doi.org/10.1103/PhysRevLett.128.190402} {\bibfield  {journal}
  {\bibinfo  {journal} {Phys. Rev. Lett.}\ }\textbf {\bibinfo {volume} {128}},\
  \bibinfo {pages} {190402} (\bibinfo {year} {2022})}\BibitemShut {NoStop}%
\bibitem [{\citenamefont {Matsoukas-Roubeas}\ \emph
  {et~al.}(2023{\natexlab{a}})\citenamefont {Matsoukas-Roubeas}, \citenamefont
  {Roccati}, \citenamefont {Cornelius}, \citenamefont {Xu}, \citenamefont
  {Chenu},\ and\ \citenamefont {del Campo}}]{Matsoukas2023}%
  \BibitemOpen
  \bibfield  {author} {\bibinfo {author} {\bibfnamefont {A.~S.}\ \bibnamefont
  {Matsoukas-Roubeas}}, \bibinfo {author} {\bibfnamefont {F.}~\bibnamefont
  {Roccati}}, \bibinfo {author} {\bibfnamefont {J.}~\bibnamefont {Cornelius}},
  \bibinfo {author} {\bibfnamefont {Z.}~\bibnamefont {Xu}}, \bibinfo {author}
  {\bibfnamefont {A.}~\bibnamefont {Chenu}},\ and\ \bibinfo {author}
  {\bibfnamefont {A.}~\bibnamefont {del Campo}},\ }\bibfield  {title} {\bibinfo
  {title} {Non-{H}ermitian {H}amiltonian deformations in quantum mechanics},\
  }\href {https://doi.org/10.1007/JHEP01(2023)060} {\bibfield  {journal}
  {\bibinfo  {journal} {J. High Energy Phys.}\ }\textbf {\bibinfo {volume}
  {2023}}\bibinfo  {number} { (1)},\ \bibinfo {pages} {60}}\BibitemShut
  {NoStop}%
\bibitem [{\citenamefont {Tameshtit}\ and\ \citenamefont
  {Sipe}(1992)}]{Tameshtit1992}%
  \BibitemOpen
\bibfield  {number} {  }\bibfield  {author} {\bibinfo {author} {\bibfnamefont
  {A.}~\bibnamefont {Tameshtit}}\ and\ \bibinfo {author} {\bibfnamefont
  {J.~E.}\ \bibnamefont {Sipe}},\ }\bibfield  {title} {\bibinfo {title}
  {Survival probability and chaos in an open quantum system},\ }\href
  {https://doi.org/10.1103/PhysRevA.45.8280} {\bibfield  {journal} {\bibinfo
  {journal} {Phys. Rev. A}\ }\textbf {\bibinfo {volume} {45}},\ \bibinfo
  {pages} {8280} (\bibinfo {year} {1992})}\BibitemShut {NoStop}%
\bibitem [{\citenamefont {{del Campo}}\ and\ \citenamefont
  {{Takayanagi}}(2020)}]{delcampo19}%
  \BibitemOpen
  \bibfield  {author} {\bibinfo {author} {\bibfnamefont {A.}~\bibnamefont {{del
  Campo}}}\ and\ \bibinfo {author} {\bibfnamefont {T.}~\bibnamefont
  {{Takayanagi}}},\ }\bibfield  {title} {\bibinfo {title} {{Decoherence in
  Conformal Field Theory}},\ }\href {https://doi.org/10.1007/JHEP02(2020)170}
  {\bibfield  {journal} {\bibinfo  {journal} {JHEP}\ }\textbf {\bibinfo
  {volume} {2020}}\bibinfo  {number} { (170)},\ \bibinfo {pages}
  {170}}\BibitemShut {NoStop}%
\bibitem [{\citenamefont {Xu}\ \emph {et~al.}(2021)\citenamefont {Xu},
  \citenamefont {Chenu}, \citenamefont {Prosen},\ and\ \citenamefont {del
  Campo}}]{Xu2021}%
  \BibitemOpen
\bibfield  {number} {  }\bibfield  {author} {\bibinfo {author} {\bibfnamefont
  {Z.}~\bibnamefont {Xu}}, \bibinfo {author} {\bibfnamefont {A.}~\bibnamefont
  {Chenu}}, \bibinfo {author} {\bibfnamefont {T.}~\bibnamefont {Prosen}},\ and\
  \bibinfo {author} {\bibfnamefont {A.}~\bibnamefont {del Campo}},\ }\bibfield
  {title} {\bibinfo {title} {Thermofield dynamics: Quantum chaos versus
  decoherence},\ }\href {https://doi.org/10.1103/PhysRevB.103.064309}
  {\bibfield  {journal} {\bibinfo  {journal} {Phys. Rev. B}\ }\textbf {\bibinfo
  {volume} {103}},\ \bibinfo {pages} {064309} (\bibinfo {year}
  {2021})}\BibitemShut {NoStop}%
\bibitem [{\citenamefont {Matsoukas-Roubeas}\ \emph
  {et~al.}(2023{\natexlab{b}})\citenamefont {Matsoukas-Roubeas}, \citenamefont
  {Prosen},\ and\ \citenamefont {del Campo}}]{Matsoukas2023quantum}%
  \BibitemOpen
  \bibfield  {author} {\bibinfo {author} {\bibfnamefont {A.~S.}\ \bibnamefont
  {Matsoukas-Roubeas}}, \bibinfo {author} {\bibfnamefont {T.}~\bibnamefont
  {Prosen}},\ and\ \bibinfo {author} {\bibfnamefont {A.}~\bibnamefont {del
  Campo}},\ }\href@noop {} {\bibinfo {title} {Quantum chaos and coherence:
  Random parametric quantum channels}} (\bibinfo {year} {2023}{\natexlab{b}}),\
  \Eprint {https://arxiv.org/abs/2305.19326} {arXiv:2305.19326 [quant-ph]}
  \BibitemShut {NoStop}%
\bibitem [{\citenamefont {Matsoukas-Roubeas}\ \emph
  {et~al.}(2023{\natexlab{c}})\citenamefont {Matsoukas-Roubeas}, \citenamefont
  {Beau}, \citenamefont {Santos},\ and\ \citenamefont {del
  Campo}}]{Matsoukas2023pra}%
  \BibitemOpen
  \bibfield  {author} {\bibinfo {author} {\bibfnamefont {A.~S.}\ \bibnamefont
  {Matsoukas-Roubeas}}, \bibinfo {author} {\bibfnamefont {M.}~\bibnamefont
  {Beau}}, \bibinfo {author} {\bibfnamefont {L.~F.}\ \bibnamefont {Santos}},\
  and\ \bibinfo {author} {\bibfnamefont {A.}~\bibnamefont {del Campo}},\
  }\bibfield  {title} {\bibinfo {title} {Unitarity breaking in self-averaging
  spectral form factors},\ }\href {https://doi.org/10.1103/PhysRevA.108.062201}
  {\bibfield  {journal} {\bibinfo  {journal} {Phys. Rev. A}\ }\textbf {\bibinfo
  {volume} {108}},\ \bibinfo {pages} {062201} (\bibinfo {year}
  {2023}{\natexlab{c}})}\BibitemShut {NoStop}%
\bibitem [{\citenamefont {Wiseman}\ and\ \citenamefont
  {Domany}(1995)}]{Wiseman1995}%
  \BibitemOpen
  \bibfield  {author} {\bibinfo {author} {\bibfnamefont {S.}~\bibnamefont
  {Wiseman}}\ and\ \bibinfo {author} {\bibfnamefont {E.}~\bibnamefont
  {Domany}},\ }\bibfield  {title} {\bibinfo {title} {Lack of self-averaging in
  critical disordered systems},\ }\href
  {https://doi.org/10.1103/PhysRevE.52.3469} {\bibfield  {journal} {\bibinfo
  {journal} {Phys. Rev. E}\ }\textbf {\bibinfo {volume} {52}},\ \bibinfo
  {pages} {3469} (\bibinfo {year} {1995})}\BibitemShut {NoStop}%
\bibitem [{\citenamefont {Aharony}\ and\ \citenamefont
  {Harris}(1996)}]{Aharony1996}%
  \BibitemOpen
  \bibfield  {author} {\bibinfo {author} {\bibfnamefont {A.}~\bibnamefont
  {Aharony}}\ and\ \bibinfo {author} {\bibfnamefont {A.~B.}\ \bibnamefont
  {Harris}},\ }\bibfield  {title} {\bibinfo {title} {{A}bsence of
  {S}elf-{A}veraging and {U}niversal {F}luctuations in {R}andom {S}ystems near
  {C}ritical {P}oints},\ }\href {https://doi.org/10.1103/PhysRevLett.77.3700}
  {\bibfield  {journal} {\bibinfo  {journal} {Phys. Rev. Lett.}\ }\textbf
  {\bibinfo {volume} {77}},\ \bibinfo {pages} {3700} (\bibinfo {year}
  {1996})}\BibitemShut {NoStop}%
\bibitem [{\citenamefont {Wiseman}\ and\ \citenamefont
  {Domany}(1998)}]{Wiseman1998}%
  \BibitemOpen
  \bibfield  {author} {\bibinfo {author} {\bibfnamefont {S.}~\bibnamefont
  {Wiseman}}\ and\ \bibinfo {author} {\bibfnamefont {E.}~\bibnamefont
  {Domany}},\ }\bibfield  {title} {\bibinfo {title} {{F}inite-{S}ize {S}caling
  and {L}ack of {S}elf-{A}veraging in {C}ritical {D}isordered {S}ystems},\
  }\href {https://doi.org/10.1103/PhysRevLett.81.22} {\bibfield  {journal}
  {\bibinfo  {journal} {Phys. Rev. Lett.}\ }\textbf {\bibinfo {volume} {81}},\
  \bibinfo {pages} {22} (\bibinfo {year} {1998})}\BibitemShut {NoStop}%
\bibitem [{\citenamefont {Castellani}\ and\ \citenamefont
  {Cavagna}(2005)}]{Castellani2005}%
  \BibitemOpen
  \bibfield  {author} {\bibinfo {author} {\bibfnamefont {T.}~\bibnamefont
  {Castellani}}\ and\ \bibinfo {author} {\bibfnamefont {A.}~\bibnamefont
  {Cavagna}},\ }\bibfield  {title} {\bibinfo {title} {Spin-glass theory for
  pedestrians},\ }\href {https://doi.org/10.1088/1742-5468/2005/05/p05012}
  {\bibfield  {journal} {\bibinfo  {journal} {J. Stat. Mech. Th. Exp.}\
  }\textbf {\bibinfo {volume} {2005}},\ \bibinfo {pages} {P05012} (\bibinfo
  {year} {2005})}\BibitemShut {NoStop}%
\bibitem [{\citenamefont {Malakis}\ and\ \citenamefont
  {Fytas}(2006)}]{Malakis2006}%
  \BibitemOpen
  \bibfield  {author} {\bibinfo {author} {\bibfnamefont {A.}~\bibnamefont
  {Malakis}}\ and\ \bibinfo {author} {\bibfnamefont {N.~G.}\ \bibnamefont
  {Fytas}},\ }\bibfield  {title} {\bibinfo {title} {Lack of self-averaging of
  the specific heat in the three-dimensional random-field {I}sing model},\
  }\href {https://doi.org/10.1103/PhysRevE.73.016109} {\bibfield  {journal}
  {\bibinfo  {journal} {Phys. Rev. E}\ }\textbf {\bibinfo {volume} {73}},\
  \bibinfo {pages} {016109} (\bibinfo {year} {2006})}\BibitemShut {NoStop}%
\bibitem [{\citenamefont {Roy}\ and\ \citenamefont
  {Bhattacharjee}(2006)}]{Roy2006}%
  \BibitemOpen
  \bibfield  {author} {\bibinfo {author} {\bibfnamefont {S.}~\bibnamefont
  {Roy}}\ and\ \bibinfo {author} {\bibfnamefont {S.~M.}\ \bibnamefont
  {Bhattacharjee}},\ }\bibfield  {title} {\bibinfo {title} {Is small-world
  network disordered?},\ }\href
  {https://doi.org/https://doi.org/10.1016/j.physleta.2005.10.105} {\bibfield
  {journal} {\bibinfo  {journal} {Phys. Lett. A}\ }\textbf {\bibinfo {volume}
  {352}},\ \bibinfo {pages} {13} (\bibinfo {year} {2006})}\BibitemShut
  {NoStop}%
\bibitem [{\citenamefont {Monthus}(2006)}]{Monthus2006}%
  \BibitemOpen
  \bibfield  {author} {\bibinfo {author} {\bibfnamefont {C.}~\bibnamefont
  {Monthus}},\ }\bibfield  {title} {\bibinfo {title} {{R}andom {W}alks and
  {P}olymers in the {P}resence of {Q}uenched {D}isorder},\ }\href
  {https://doi.org/10.1007/s11005-006-0122-2} {\bibfield  {journal} {\bibinfo
  {journal} {Lett. Math. Phys.}\ }\textbf {\bibinfo {volume} {78}},\ \bibinfo
  {pages} {207} (\bibinfo {year} {2006})}\BibitemShut {NoStop}%
\bibitem [{\citenamefont {Efrat}\ and\ \citenamefont
  {Schwartz}(2014)}]{Efrat2014}%
  \BibitemOpen
  \bibfield  {author} {\bibinfo {author} {\bibfnamefont {A.}~\bibnamefont
  {Efrat}}\ and\ \bibinfo {author} {\bibfnamefont {M.}~\bibnamefont
  {Schwartz}},\ }\bibfield  {title} {\bibinfo {title} {{L}ack of self-averaging
  in random systems - {L}iability or asset?},\ }\href
  {https://doi.org/https://doi.org/10.1016/j.physa.2014.06.071} {\bibfield
  {journal} {\bibinfo  {journal} {Phys. A Stat. Mech. Appl.}\ }\textbf
  {\bibinfo {volume} {414}},\ \bibinfo {pages} {137} (\bibinfo {year}
  {2014})}\BibitemShut {NoStop}%
\bibitem [{\citenamefont {\L{}obejko}\ \emph {et~al.}(2018)\citenamefont
  {\L{}obejko}, \citenamefont {Dajka},\ and\ \citenamefont
  {\L{}uczka}}]{Lobejko2018}%
  \BibitemOpen
  \bibfield  {author} {\bibinfo {author} {\bibfnamefont {M.}~\bibnamefont
  {\L{}obejko}}, \bibinfo {author} {\bibfnamefont {J.}~\bibnamefont {Dajka}},\
  and\ \bibinfo {author} {\bibfnamefont {J.}~\bibnamefont {\L{}uczka}},\
  }\bibfield  {title} {\bibinfo {title} {Self-averaging of random quantum
  dynamics},\ }\href {https://doi.org/10.1103/PhysRevA.98.022111} {\bibfield
  {journal} {\bibinfo  {journal} {Phys. Rev. A}\ }\textbf {\bibinfo {volume}
  {98}},\ \bibinfo {pages} {022111} (\bibinfo {year} {2018})}\BibitemShut
  {NoStop}%
\bibitem [{\citenamefont {Gredeskul}\ and\ \citenamefont
  {Pastur}(1985)}]{Gredeskul1985}%
  \BibitemOpen
  \bibfield  {author} {\bibinfo {author} {\bibfnamefont {S.~A.}\ \bibnamefont
  {Gredeskul}}\ and\ \bibinfo {author} {\bibfnamefont {L.~A.}\ \bibnamefont
  {Pastur}},\ }\bibfield  {title} {\bibinfo {title} {Works of {I.} {M.}
  {L}ifshitz on disordered systems},\ }\href
  {https://doi.org/10.1007/BF01017846} {\bibfield  {journal} {\bibinfo
  {journal} {J. Stat. Phys.}\ }\textbf {\bibinfo {volume} {38}},\ \bibinfo
  {pages} {25} (\bibinfo {year} {1985})}\BibitemShut {NoStop}%
\bibitem [{\citenamefont {M\"uller}\ and\ \citenamefont
  {Delande}(2016)}]{MullerARXIV}%
  \BibitemOpen
  \bibfield  {author} {\bibinfo {author} {\bibfnamefont {C.~A.}\ \bibnamefont
  {M\"uller}}\ and\ \bibinfo {author} {\bibfnamefont {D.}~\bibnamefont
  {Delande}},\ }\href@noop {} {\bibinfo {title} {{D}isorder and interference:
  localization phenomena}} (\bibinfo {year} {2016}),\ \bibinfo {note} {les
  Houches 2009 - Session XCI: Ultracold Gases and Quantum Information, C.
  Miniatura, L.-C. Kwek, M. Ducloy, B. Gremaud, B.-G. Englert, L.F.
  Cugliandolo, A. Ekert, eds. (Oxford University Press, Oxford 2011);
  arXiv:1004.0915},\ \Eprint {https://arxiv.org/abs/1005.0915} {arXiv:1005.0915
  [cond-mat.dis-nn]} \BibitemShut {NoStop}%
\bibitem [{\citenamefont {Serbyn}\ \emph {et~al.}(2017)\citenamefont {Serbyn},
  \citenamefont {Papi\ifmmode~\acute{c}\else \'{c}\fi{}},\ and\ \citenamefont
  {Abanin}}]{Serbyn2017}%
  \BibitemOpen
  \bibfield  {author} {\bibinfo {author} {\bibfnamefont {M.}~\bibnamefont
  {Serbyn}}, \bibinfo {author} {\bibfnamefont {Z.}~\bibnamefont
  {Papi\ifmmode~\acute{c}\else \'{c}\fi{}}},\ and\ \bibinfo {author}
  {\bibfnamefont {D.~A.}\ \bibnamefont {Abanin}},\ }\bibfield  {title}
  {\bibinfo {title} {Thouless energy and multifractality across the many-body
  localization transition},\ }\href
  {https://doi.org/10.1103/PhysRevB.96.104201} {\bibfield  {journal} {\bibinfo
  {journal} {Phys. Rev. B}\ }\textbf {\bibinfo {volume} {96}},\ \bibinfo
  {pages} {104201} (\bibinfo {year} {2017})}\BibitemShut {NoStop}%
\bibitem [{\citenamefont {Pastur}\ and\ \citenamefont
  {Slavin}(2014)}]{Pastur2014PRL}%
  \BibitemOpen
  \bibfield  {author} {\bibinfo {author} {\bibfnamefont {L.}~\bibnamefont
  {Pastur}}\ and\ \bibinfo {author} {\bibfnamefont {V.}~\bibnamefont
  {Slavin}},\ }\bibfield  {title} {\bibinfo {title} {{A}rea {L}aw {S}caling for
  the {E}ntropy of {D}isordered {Q}uasifree {F}ermions},\ }\href
  {https://doi.org/10.1103/PhysRevLett.113.150404} {\bibfield  {journal}
  {\bibinfo  {journal} {Phys. Rev. Lett.}\ }\textbf {\bibinfo {volume} {113}},\
  \bibinfo {pages} {150404} (\bibinfo {year} {2014})}\BibitemShut {NoStop}%
\bibitem [{\citenamefont {Milchev}\ \emph {et~al.}(1986)\citenamefont
  {Milchev}, \citenamefont {Binder},\ and\ \citenamefont
  {Heermann}}]{Milchev1986}%
  \BibitemOpen
  \bibfield  {author} {\bibinfo {author} {\bibfnamefont {A.}~\bibnamefont
  {Milchev}}, \bibinfo {author} {\bibfnamefont {K.}~\bibnamefont {Binder}},\
  and\ \bibinfo {author} {\bibfnamefont {D.~W.}\ \bibnamefont {Heermann}},\
  }\bibfield  {title} {\bibinfo {title} {Fluctuations and lack of
  self-averaging in the kinetics of domain growth},\ }\href
  {https://doi.org/10.1007/BF01726202} {\bibfield  {journal} {\bibinfo
  {journal} {Z. Phys. B Condensed Matter}\ }\textbf {\bibinfo {volume} {63}},\
  \bibinfo {pages} {521} (\bibinfo {year} {1986})}\BibitemShut {NoStop}%
\bibitem [{\citenamefont {Bouchaud}\ and\ \citenamefont
  {Georges}(1990)}]{Bouchaud1990}%
  \BibitemOpen
  \bibfield  {author} {\bibinfo {author} {\bibfnamefont {J.-P.}\ \bibnamefont
  {Bouchaud}}\ and\ \bibinfo {author} {\bibfnamefont {A.}~\bibnamefont
  {Georges}},\ }\bibfield  {title} {\bibinfo {title} {Anomalous diffusion in
  disordered media: {S}tatistical mechanisms, models and physical
  applications},\ }\href
  {https://doi.org/https://doi.org/10.1016/0370-1573(90)90099-N} {\bibfield
  {journal} {\bibinfo  {journal} {Phys. Rep.}\ }\textbf {\bibinfo {volume}
  {195}},\ \bibinfo {pages} {127} (\bibinfo {year} {1990})}\BibitemShut
  {NoStop}%
\bibitem [{\citenamefont {Akimoto}\ \emph {et~al.}(2016)\citenamefont
  {Akimoto}, \citenamefont {Barkai},\ and\ \citenamefont
  {Saito}}]{AkimotoPRL2016}%
  \BibitemOpen
  \bibfield  {author} {\bibinfo {author} {\bibfnamefont {T.}~\bibnamefont
  {Akimoto}}, \bibinfo {author} {\bibfnamefont {E.}~\bibnamefont {Barkai}},\
  and\ \bibinfo {author} {\bibfnamefont {K.}~\bibnamefont {Saito}},\ }\bibfield
   {title} {\bibinfo {title} {{U}niversal {F}luctuations of {S}ingle-{P}article
  {D}iffusivity in a {Q}uenched {E}nvironment},\ }\href
  {https://doi.org/10.1103/PhysRevLett.117.180602} {\bibfield  {journal}
  {\bibinfo  {journal} {Phys. Rev. Lett.}\ }\textbf {\bibinfo {volume} {117}},\
  \bibinfo {pages} {180602} (\bibinfo {year} {2016})}\BibitemShut {NoStop}%
\bibitem [{\citenamefont {Russian}\ \emph {et~al.}(2017)\citenamefont
  {Russian}, \citenamefont {Dentz},\ and\ \citenamefont {Gouze}}]{Russian2017}%
  \BibitemOpen
  \bibfield  {author} {\bibinfo {author} {\bibfnamefont {A.}~\bibnamefont
  {Russian}}, \bibinfo {author} {\bibfnamefont {M.}~\bibnamefont {Dentz}},\
  and\ \bibinfo {author} {\bibfnamefont {P.}~\bibnamefont {Gouze}},\ }\bibfield
   {title} {\bibinfo {title} {Self-averaging and weak ergodicity breaking of
  diffusion in heterogeneous media},\ }\href
  {https://doi.org/10.1103/PhysRevE.96.022156} {\bibfield  {journal} {\bibinfo
  {journal} {Phys. Rev. E}\ }\textbf {\bibinfo {volume} {96}},\ \bibinfo
  {pages} {022156} (\bibinfo {year} {2017})}\BibitemShut {NoStop}%
\bibitem [{\citenamefont {Akimoto}\ \emph {et~al.}(2018)\citenamefont
  {Akimoto}, \citenamefont {Barkai},\ and\ \citenamefont
  {Saito}}]{AkimotoPRE2018}%
  \BibitemOpen
  \bibfield  {author} {\bibinfo {author} {\bibfnamefont {T.}~\bibnamefont
  {Akimoto}}, \bibinfo {author} {\bibfnamefont {E.}~\bibnamefont {Barkai}},\
  and\ \bibinfo {author} {\bibfnamefont {K.}~\bibnamefont {Saito}},\ }\bibfield
   {title} {\bibinfo {title} {Non-self-averaging behaviors and ergodicity in
  quenched trap models with finite system sizes},\ }\href
  {https://doi.org/10.1103/PhysRevE.97.052143} {\bibfield  {journal} {\bibinfo
  {journal} {Phys. Rev. E}\ }\textbf {\bibinfo {volume} {97}},\ \bibinfo
  {pages} {052143} (\bibinfo {year} {2018})}\BibitemShut {NoStop}%
\bibitem [{\citenamefont {Wreszinski}\ and\ \citenamefont
  {Bolina}(2004)}]{Wreszinski2004}%
  \BibitemOpen
  \bibfield  {author} {\bibinfo {author} {\bibfnamefont {W.}~\bibnamefont
  {Wreszinski}}\ and\ \bibinfo {author} {\bibfnamefont {O.}~\bibnamefont
  {Bolina}},\ }\bibfield  {title} {\bibinfo {title} {A self-averaging ``order
  parameter'' for the {S}herrington-{K}irkpatrick spin glass model},\ }\href
  {https://doi.org/10.1023/B:JOSS.0000041743.24497.63} {\bibfield  {journal}
  {\bibinfo  {journal} {J. Stat. Phys.}\ }\textbf {\bibinfo {volume} {116}},\
  \bibinfo {pages} {1389} (\bibinfo {year} {2004})}\BibitemShut {NoStop}%
\bibitem [{\citenamefont {Sol\'orzano}\ \emph {et~al.}(2021)\citenamefont
  {Sol\'orzano}, \citenamefont {Santos},\ and\ \citenamefont
  {Torres-Herrera}}]{Solorzano2021}%
  \BibitemOpen
  \bibfield  {author} {\bibinfo {author} {\bibfnamefont {A.}~\bibnamefont
  {Sol\'orzano}}, \bibinfo {author} {\bibfnamefont {L.~F.}\ \bibnamefont
  {Santos}},\ and\ \bibinfo {author} {\bibfnamefont {E.~J.}\ \bibnamefont
  {Torres-Herrera}},\ }\bibfield  {title} {\bibinfo {title} {Multifractality
  and self-averaging at the many-body localization transition},\ }\href
  {https://doi.org/10.1103/PhysRevResearch.3.L032030} {\bibfield  {journal}
  {\bibinfo  {journal} {Phys. Rev. Res.}\ }\textbf {\bibinfo {volume} {3}},\
  \bibinfo {pages} {L032030} (\bibinfo {year} {2021})}\BibitemShut {NoStop}%
\bibitem [{\citenamefont {Ithier}\ and\ \citenamefont
  {Benaych-Georges}(2017)}]{Ithier2017}%
  \BibitemOpen
  \bibfield  {author} {\bibinfo {author} {\bibfnamefont {G.}~\bibnamefont
  {Ithier}}\ and\ \bibinfo {author} {\bibfnamefont {F.}~\bibnamefont
  {Benaych-Georges}},\ }\bibfield  {title} {\bibinfo {title} {Dynamical
  typicality of embedded quantum systems},\ }\href
  {https://doi.org/10.1103/PhysRevA.96.012108} {\bibfield  {journal} {\bibinfo
  {journal} {Phys. Rev. A}\ }\textbf {\bibinfo {volume} {96}},\ \bibinfo
  {pages} {012108} (\bibinfo {year} {2017})}\BibitemShut {NoStop}%
\bibitem [{\citenamefont {Mukherjee}(2018)}]{Mukherjee2018}%
  \BibitemOpen
  \bibfield  {author} {\bibinfo {author} {\bibfnamefont {B.}~\bibnamefont
  {Mukherjee}},\ }\bibfield  {title} {\bibinfo {title} {Floquet topological
  transition by unpolarized light},\ }\href
  {https://doi.org/10.1103/PhysRevB.98.235112} {\bibfield  {journal} {\bibinfo
  {journal} {Phys. Rev. B}\ }\textbf {\bibinfo {volume} {98}},\ \bibinfo
  {pages} {235112} (\bibinfo {year} {2018})}\BibitemShut {NoStop}%
\bibitem [{\citenamefont {Richter}\ \emph {et~al.}(2020)\citenamefont
  {Richter}, \citenamefont {Schubert},\ and\ \citenamefont
  {Steinigeweg}}]{Richter2020}%
  \BibitemOpen
  \bibfield  {author} {\bibinfo {author} {\bibfnamefont {J.}~\bibnamefont
  {Richter}}, \bibinfo {author} {\bibfnamefont {D.}~\bibnamefont {Schubert}},\
  and\ \bibinfo {author} {\bibfnamefont {R.}~\bibnamefont {Steinigeweg}},\
  }\bibfield  {title} {\bibinfo {title} {Decay of spin-spin correlations in
  disordered quantum and classical spin chains},\ }\href
  {https://doi.org/10.1103/PhysRevResearch.2.013130} {\bibfield  {journal}
  {\bibinfo  {journal} {Phys. Rev. Res.}\ }\textbf {\bibinfo {volume} {2}},\
  \bibinfo {pages} {013130} (\bibinfo {year} {2020})}\BibitemShut {NoStop}%
\bibitem [{\citenamefont {Schiulaz}\ \emph {et~al.}(2020)\citenamefont
  {Schiulaz}, \citenamefont {Torres-Herrera}, \citenamefont {P\'erez-Bernal},\
  and\ \citenamefont {Santos}}]{Schiulaz2020}%
  \BibitemOpen
  \bibfield  {author} {\bibinfo {author} {\bibfnamefont {M.}~\bibnamefont
  {Schiulaz}}, \bibinfo {author} {\bibfnamefont {E.~J.}\ \bibnamefont
  {Torres-Herrera}}, \bibinfo {author} {\bibfnamefont {F.}~\bibnamefont
  {P\'erez-Bernal}},\ and\ \bibinfo {author} {\bibfnamefont {L.~F.}\
  \bibnamefont {Santos}},\ }\bibfield  {title} {\bibinfo {title}
  {{S}elf-averaging in many-body quantum systems out of equilibrium: {C}haotic
  systems},\ }\href {https://doi.org/10.1103/PhysRevB.101.174312} {\bibfield
  {journal} {\bibinfo  {journal} {Phys. Rev. B}\ }\textbf {\bibinfo {volume}
  {101}},\ \bibinfo {pages} {174312} (\bibinfo {year} {2020})}\BibitemShut
  {NoStop}%
\bibitem [{\citenamefont {Torres-Herrera}\ \emph
  {et~al.}(2020{\natexlab{a}})\citenamefont {Torres-Herrera}, \citenamefont
  {Vallejo-Fabila}, \citenamefont {Mart\'{\i}nez-Mendoza},\ and\ \citenamefont
  {Santos}}]{Torres2020PRE}%
  \BibitemOpen
  \bibfield  {author} {\bibinfo {author} {\bibfnamefont {E.~J.}\ \bibnamefont
  {Torres-Herrera}}, \bibinfo {author} {\bibfnamefont {I.}~\bibnamefont
  {Vallejo-Fabila}}, \bibinfo {author} {\bibfnamefont {A.~J.}\ \bibnamefont
  {Mart\'{\i}nez-Mendoza}},\ and\ \bibinfo {author} {\bibfnamefont {L.~F.}\
  \bibnamefont {Santos}},\ }\bibfield  {title} {\bibinfo {title}
  {Self-averaging in many-body quantum systems out of equilibrium: Time
  dependence of distributions},\ }\href
  {https://doi.org/10.1103/PhysRevE.102.062126} {\bibfield  {journal} {\bibinfo
   {journal} {Phys. Rev. E}\ }\textbf {\bibinfo {volume} {102}},\ \bibinfo
  {pages} {062126} (\bibinfo {year} {2020}{\natexlab{a}})}\BibitemShut
  {NoStop}%
\bibitem [{\citenamefont {Torres-Herrera}\ \emph
  {et~al.}(2020{\natexlab{b}})\citenamefont {Torres-Herrera}, \citenamefont
  {De~Tomasi}, \citenamefont {Schiulaz}, \citenamefont {P\'erez-Bernal},\ and\
  \citenamefont {Santos}}]{Torres2020}%
  \BibitemOpen
  \bibfield  {author} {\bibinfo {author} {\bibfnamefont {E.~J.}\ \bibnamefont
  {Torres-Herrera}}, \bibinfo {author} {\bibfnamefont {G.}~\bibnamefont
  {De~Tomasi}}, \bibinfo {author} {\bibfnamefont {M.}~\bibnamefont {Schiulaz}},
  \bibinfo {author} {\bibfnamefont {F.}~\bibnamefont {P\'erez-Bernal}},\ and\
  \bibinfo {author} {\bibfnamefont {L.~F.}\ \bibnamefont {Santos}},\ }\bibfield
   {title} {\bibinfo {title} {Self-averaging in many-body quantum systems out
  of equilibrium: Approach to the localized phase},\ }\href
  {https://doi.org/10.1103/PhysRevB.102.094310} {\bibfield  {journal} {\bibinfo
   {journal} {Phys. Rev. B}\ }\textbf {\bibinfo {volume} {102}},\ \bibinfo
  {pages} {094310} (\bibinfo {year} {2020}{\natexlab{b}})}\BibitemShut
  {NoStop}%
\bibitem [{\citenamefont {Argaman}\ \emph {et~al.}(1993)\citenamefont
  {Argaman}, \citenamefont {Dittes}, \citenamefont {Doron}, \citenamefont
  {Keating}, \citenamefont {Kitaev}, \citenamefont {Sieber},\ and\
  \citenamefont {Smilansky}}]{Argaman1993b}%
  \BibitemOpen
  \bibfield  {author} {\bibinfo {author} {\bibfnamefont {N.}~\bibnamefont
  {Argaman}}, \bibinfo {author} {\bibfnamefont {F.-M.}\ \bibnamefont {Dittes}},
  \bibinfo {author} {\bibfnamefont {E.}~\bibnamefont {Doron}}, \bibinfo
  {author} {\bibfnamefont {J.~P.}\ \bibnamefont {Keating}}, \bibinfo {author}
  {\bibfnamefont {A.~Y.}\ \bibnamefont {Kitaev}}, \bibinfo {author}
  {\bibfnamefont {M.}~\bibnamefont {Sieber}},\ and\ \bibinfo {author}
  {\bibfnamefont {U.}~\bibnamefont {Smilansky}},\ }\bibfield  {title} {\bibinfo
  {title} {Correlations in the actions of periodic orbits derived from quantum
  chaos},\ }\href {https://doi.org/10.1103/PhysRevLett.71.4326} {\bibfield
  {journal} {\bibinfo  {journal} {Phys. Rev. Lett.}\ }\textbf {\bibinfo
  {volume} {71}},\ \bibinfo {pages} {4326} (\bibinfo {year}
  {1993})}\BibitemShut {NoStop}%
\bibitem [{\citenamefont {Eckhardt}\ and\ \citenamefont
  {Main}(1995)}]{Eckhardt1995}%
  \BibitemOpen
  \bibfield  {author} {\bibinfo {author} {\bibfnamefont {B.}~\bibnamefont
  {Eckhardt}}\ and\ \bibinfo {author} {\bibfnamefont {J.}~\bibnamefont
  {Main}},\ }\bibfield  {title} {\bibinfo {title} {Semiclassical form factor of
  matrix element fluctuations},\ }\href
  {https://doi.org/10.1103/PhysRevLett.75.2300} {\bibfield  {journal} {\bibinfo
   {journal} {Phys. Rev. Lett.}\ }\textbf {\bibinfo {volume} {75}},\ \bibinfo
  {pages} {2300} (\bibinfo {year} {1995})}\BibitemShut {NoStop}%
\bibitem [{\citenamefont {Prange}(1997)}]{Prange1997}%
  \BibitemOpen
  \bibfield  {author} {\bibinfo {author} {\bibfnamefont {R.~E.}\ \bibnamefont
  {Prange}},\ }\bibfield  {title} {\bibinfo {title} {The spectral form factor
  is not self-averaging},\ }\href {https://doi.org/10.1103/PhysRevLett.78.2280}
  {\bibfield  {journal} {\bibinfo  {journal} {Phys. Rev. Lett.}\ }\textbf
  {\bibinfo {volume} {78}},\ \bibinfo {pages} {2280} (\bibinfo {year}
  {1997})}\BibitemShut {NoStop}%
\bibitem [{\citenamefont {Braun}\ and\ \citenamefont
  {Haake}(2015)}]{Braun2015}%
  \BibitemOpen
  \bibfield  {author} {\bibinfo {author} {\bibfnamefont {P.}~\bibnamefont
  {Braun}}\ and\ \bibinfo {author} {\bibfnamefont {F.}~\bibnamefont {Haake}},\
  }\bibfield  {title} {\bibinfo {title} {Self-averaging characteristics of
  spectral fluctuations},\ }\href
  {https://doi.org/10.1088/1751-8113/48/13/135101} {\bibfield  {journal}
  {\bibinfo  {journal} {J.Phys. A}\ }\textbf {\bibinfo {volume} {48}},\
  \bibinfo {pages} {135101} (\bibinfo {year} {2015})}\BibitemShut {NoStop}%
\bibitem [{\citenamefont {Mehta}(1991)}]{MehtaBook}%
  \BibitemOpen
  \bibfield  {author} {\bibinfo {author} {\bibfnamefont {M.~L.}\ \bibnamefont
  {Mehta}},\ }\href@noop {} {\emph {\bibinfo {title} {Random Matrices}}}\
  (\bibinfo  {publisher} {Academic Press},\ \bibinfo {address} {Boston},\
  \bibinfo {year} {1991})\BibitemShut {NoStop}%
\bibitem [{\citenamefont {Leviandier}\ \emph {et~al.}(1986)\citenamefont
  {Leviandier}, \citenamefont {Lombardi}, \citenamefont {Jost},\ and\
  \citenamefont {Pique}}]{Leviandier1986}%
  \BibitemOpen
  \bibfield  {author} {\bibinfo {author} {\bibfnamefont {L.}~\bibnamefont
  {Leviandier}}, \bibinfo {author} {\bibfnamefont {M.}~\bibnamefont
  {Lombardi}}, \bibinfo {author} {\bibfnamefont {R.}~\bibnamefont {Jost}},\
  and\ \bibinfo {author} {\bibfnamefont {J.~P.}\ \bibnamefont {Pique}},\
  }\bibfield  {title} {\bibinfo {title} {Fourier transform: A tool to measure
  statistical level properties in very complex spectra},\ }\href
  {https://doi.org/10.1103/PhysRevLett.56.2449} {\bibfield  {journal} {\bibinfo
   {journal} {Phys. Rev. Lett.}\ }\textbf {\bibinfo {volume} {56}},\ \bibinfo
  {pages} {2449} (\bibinfo {year} {1986})}\BibitemShut {NoStop}%
\bibitem [{\citenamefont {Pique}\ \emph {et~al.}(1987)\citenamefont {Pique},
  \citenamefont {Chen}, \citenamefont {Field},\ and\ \citenamefont
  {Kinsey}}]{Pique1987}%
  \BibitemOpen
  \bibfield  {author} {\bibinfo {author} {\bibfnamefont {J.~P.}\ \bibnamefont
  {Pique}}, \bibinfo {author} {\bibfnamefont {Y.}~\bibnamefont {Chen}},
  \bibinfo {author} {\bibfnamefont {R.~W.}\ \bibnamefont {Field}},\ and\
  \bibinfo {author} {\bibfnamefont {J.~L.}\ \bibnamefont {Kinsey}},\ }\bibfield
   {title} {\bibinfo {title} {Chaos and dynamics on 0.5 -- 300 ps time scales
  in vibrationally excited acetylene: Fourier transform of stimulated-emission
  pumping spectrum},\ }\href {https://doi.org/10.1103/PhysRevLett.58.475}
  {\bibfield  {journal} {\bibinfo  {journal} {Phys. Rev. Lett.}\ }\textbf
  {\bibinfo {volume} {58}},\ \bibinfo {pages} {475} (\bibinfo {year}
  {1987})}\BibitemShut {NoStop}%
\bibitem [{\citenamefont {Guhr}\ and\ \citenamefont
  {Weidenm\"uller}(1990)}]{Guhr1990}%
  \BibitemOpen
  \bibfield  {author} {\bibinfo {author} {\bibfnamefont {T.}~\bibnamefont
  {Guhr}}\ and\ \bibinfo {author} {\bibfnamefont {H.}~\bibnamefont
  {Weidenm\"uller}},\ }\bibfield  {title} {\bibinfo {title} {Correlations in
  anticrossing spectra and scattering theory. analytical aspects},\ }\href
  {https://doi.org/http://dx.doi.org/10.1016/0301-0104(90)90003-R} {\bibfield
  {journal} {\bibinfo  {journal} {Chem. Phys.}\ }\textbf {\bibinfo {volume}
  {146}},\ \bibinfo {pages} {21 } (\bibinfo {year} {1990})}\BibitemShut
  {NoStop}%
\bibitem [{\citenamefont {Hartmann}\ \emph {et~al.}(1991)\citenamefont
  {Hartmann}, \citenamefont {Weidenm\"uller},\ and\ \citenamefont
  {Guhr}}]{Hartmann1991}%
  \BibitemOpen
  \bibfield  {author} {\bibinfo {author} {\bibfnamefont {U.}~\bibnamefont
  {Hartmann}}, \bibinfo {author} {\bibfnamefont {H.}~\bibnamefont
  {Weidenm\"uller}},\ and\ \bibinfo {author} {\bibfnamefont {T.}~\bibnamefont
  {Guhr}},\ }\bibfield  {title} {\bibinfo {title} {Correlations in anticrossing
  spectra and scattering theory: Numerical simulations},\ }\href
  {https://doi.org/http://dx.doi.org/10.1016/0301-0104(91)87105-5} {\bibfield
  {journal} {\bibinfo  {journal} {Chem. Phys.}\ }\textbf {\bibinfo {volume}
  {150}},\ \bibinfo {pages} {311 } (\bibinfo {year} {1991})}\BibitemShut
  {NoStop}%
\bibitem [{\citenamefont {Alhassid}\ and\ \citenamefont
  {Levine}(1992)}]{Alhassid1992}%
  \BibitemOpen
  \bibfield  {author} {\bibinfo {author} {\bibfnamefont {Y.}~\bibnamefont
  {Alhassid}}\ and\ \bibinfo {author} {\bibfnamefont {R.~D.}\ \bibnamefont
  {Levine}},\ }\bibfield  {title} {\bibinfo {title} {Spectral autocorrelation
  function in the statistical theory of energy levels},\ }\href
  {https://doi.org/10.1103/PhysRevA.46.4650} {\bibfield  {journal} {\bibinfo
  {journal} {Phys. Rev. A}\ }\textbf {\bibinfo {volume} {46}},\ \bibinfo
  {pages} {4650} (\bibinfo {year} {1992})}\BibitemShut {NoStop}%
\bibitem [{\citenamefont {Lombardi}\ and\ \citenamefont
  {Seligman}(1993)}]{Lombardi1993}%
  \BibitemOpen
  \bibfield  {author} {\bibinfo {author} {\bibfnamefont {M.}~\bibnamefont
  {Lombardi}}\ and\ \bibinfo {author} {\bibfnamefont {T.~H.}\ \bibnamefont
  {Seligman}},\ }\bibfield  {title} {\bibinfo {title} {Universal and
  nonuniversal statistical properties of levels and intensities for chaotic
  {R}ydberg molecules},\ }\href {https://doi.org/10.1103/PhysRevA.47.3571}
  {\bibfield  {journal} {\bibinfo  {journal} {Phys. Rev. A}\ }\textbf {\bibinfo
  {volume} {47}},\ \bibinfo {pages} {3571} (\bibinfo {year}
  {1993})}\BibitemShut {NoStop}%
\bibitem [{\citenamefont {Michaille}\ and\ \citenamefont
  {Pique}(1999)}]{Michaille1999}%
  \BibitemOpen
  \bibfield  {author} {\bibinfo {author} {\bibfnamefont {L.}~\bibnamefont
  {Michaille}}\ and\ \bibinfo {author} {\bibfnamefont {J.-P.}\ \bibnamefont
  {Pique}},\ }\bibfield  {title} {\bibinfo {title} {Influence of experimental
  resolution on the spectral statistics used to show quantum chaos: The case of
  molecular vibrational chaos},\ }\href
  {https://doi.org/10.1103/PhysRevLett.82.2083} {\bibfield  {journal} {\bibinfo
   {journal} {Phys. Rev. Lett.}\ }\textbf {\bibinfo {volume} {82}},\ \bibinfo
  {pages} {2083} (\bibinfo {year} {1999})}\BibitemShut {NoStop}%
\bibitem [{\citenamefont {Leyvraz}\ \emph {et~al.}(2013)\citenamefont
  {Leyvraz}, \citenamefont {Garc\'ia}, \citenamefont {Kohler},\ and\
  \citenamefont {Seligman}}]{Leyvraz2013}%
  \BibitemOpen
  \bibfield  {author} {\bibinfo {author} {\bibfnamefont {F.}~\bibnamefont
  {Leyvraz}}, \bibinfo {author} {\bibfnamefont {A.}~\bibnamefont {Garc\'ia}},
  \bibinfo {author} {\bibfnamefont {H.}~\bibnamefont {Kohler}},\ and\ \bibinfo
  {author} {\bibfnamefont {T.~H.}\ \bibnamefont {Seligman}},\ }\bibfield
  {title} {\bibinfo {title} {Fidelity under isospectral perturbations: a random
  matrix study},\ }\href {http://stacks.iop.org/1751-8121/46/i=27/a=275303}
  {\bibfield  {journal} {\bibinfo  {journal} {J. Phys. A}\ }\textbf {\bibinfo
  {volume} {46}},\ \bibinfo {pages} {275303} (\bibinfo {year}
  {2013})}\BibitemShut {NoStop}%
\bibitem [{\citenamefont {Torres-Herrera}\ and\ \citenamefont
  {Santos}(2017{\natexlab{a}})}]{Torres2017Philo}%
  \BibitemOpen
  \bibfield  {author} {\bibinfo {author} {\bibfnamefont {E.~J.}\ \bibnamefont
  {Torres-Herrera}}\ and\ \bibinfo {author} {\bibfnamefont {L.~F.}\
  \bibnamefont {Santos}},\ }\bibfield  {title} {\bibinfo {title} {Dynamical
  manifestations of quantum chaos: correlation hole and bulge},\ }\href
  {https://doi.org/10.1098/rsta.2016.0434} {\bibfield  {journal} {\bibinfo
  {journal} {Philos. Trans. Royal Soc. A}\ }\textbf {\bibinfo {volume} {375}},\
  \bibinfo {pages} {20160434} (\bibinfo {year}
  {2017}{\natexlab{a}})}\BibitemShut {NoStop}%
\bibitem [{\citenamefont {Torres-Herrera}\ and\ \citenamefont
  {Santos}(2017{\natexlab{b}})}]{Torres2017}%
  \BibitemOpen
  \bibfield  {author} {\bibinfo {author} {\bibfnamefont {E.~J.}\ \bibnamefont
  {Torres-Herrera}}\ and\ \bibinfo {author} {\bibfnamefont {L.~F.}\
  \bibnamefont {Santos}},\ }\bibfield  {title} {\bibinfo {title} {Extended
  nonergodic states in disordered many-body quantum systems},\ }\href
  {https://doi.org/10.1002/andp.201600284} {\bibfield  {journal} {\bibinfo
  {journal} {Ann. Phys. (Berlin)}\ }\textbf {\bibinfo {volume} {529}},\
  \bibinfo {pages} {1600284} (\bibinfo {year}
  {2017}{\natexlab{b}})}\BibitemShut {NoStop}%
\bibitem [{\citenamefont {Torres-Herrera}\ \emph {et~al.}(2018)\citenamefont
  {Torres-Herrera}, \citenamefont {Garc\'{\i}a-Garc\'{\i}a},\ and\
  \citenamefont {Santos}}]{Torres2018}%
  \BibitemOpen
  \bibfield  {author} {\bibinfo {author} {\bibfnamefont {E.~J.}\ \bibnamefont
  {Torres-Herrera}}, \bibinfo {author} {\bibfnamefont {A.~M.}\ \bibnamefont
  {Garc\'{\i}a-Garc\'{\i}a}},\ and\ \bibinfo {author} {\bibfnamefont {L.~F.}\
  \bibnamefont {Santos}},\ }\bibfield  {title} {\bibinfo {title} {Generic
  dynamical features of quenched interacting quantum systems: Survival
  probability, density imbalance, and out-of-time-ordered correlator},\ }\href
  {https://doi.org/10.1103/PhysRevB.97.060303} {\bibfield  {journal} {\bibinfo
  {journal} {Phys. Rev. B}\ }\textbf {\bibinfo {volume} {97}},\ \bibinfo
  {pages} {060303} (\bibinfo {year} {2018})}\BibitemShut {NoStop}%
\bibitem [{\citenamefont {Schiulaz}\ \emph {et~al.}(2019)\citenamefont
  {Schiulaz}, \citenamefont {Torres-Herrera},\ and\ \citenamefont
  {Santos}}]{Schiulaz2019}%
  \BibitemOpen
  \bibfield  {author} {\bibinfo {author} {\bibfnamefont {M.}~\bibnamefont
  {Schiulaz}}, \bibinfo {author} {\bibfnamefont {E.~J.}\ \bibnamefont
  {Torres-Herrera}},\ and\ \bibinfo {author} {\bibfnamefont {L.~F.}\
  \bibnamefont {Santos}},\ }\bibfield  {title} {\bibinfo {title} {{T}houless
  and relaxation time scales in many-body quantum systems},\ }\href
  {https://doi.org/10.1103/PhysRevB.99.174313} {\bibfield  {journal} {\bibinfo
  {journal} {Phys. Rev. B}\ }\textbf {\bibinfo {volume} {99}},\ \bibinfo
  {pages} {174313} (\bibinfo {year} {2019})}\BibitemShut {NoStop}%
\bibitem [{\citenamefont {Das}\ \emph {et~al.}(2024{\natexlab{a}})\citenamefont
  {Das}, \citenamefont {Cianci}, \citenamefont {Cabral}, \citenamefont
  {Zarate-Herrada}, \citenamefont {Pinney}, \citenamefont {Pilatowsky-Cameo},
  \citenamefont {Matsoukas-Roubeas}, \citenamefont {Batista}, \citenamefont
  {del Campo}, \citenamefont {Torres-Herrera},\ and\ \citenamefont
  {Santos}}]{Das2024}%
  \BibitemOpen
  \bibfield  {author} {\bibinfo {author} {\bibfnamefont {A.~K.}\ \bibnamefont
  {Das}}, \bibinfo {author} {\bibfnamefont {C.}~\bibnamefont {Cianci}},
  \bibinfo {author} {\bibfnamefont {D.~G.~A.}\ \bibnamefont {Cabral}}, \bibinfo
  {author} {\bibfnamefont {D.~A.}\ \bibnamefont {Zarate-Herrada}}, \bibinfo
  {author} {\bibfnamefont {P.}~\bibnamefont {Pinney}}, \bibinfo {author}
  {\bibfnamefont {S.}~\bibnamefont {Pilatowsky-Cameo}}, \bibinfo {author}
  {\bibfnamefont {A.~S.}\ \bibnamefont {Matsoukas-Roubeas}}, \bibinfo {author}
  {\bibfnamefont {V.~S.}\ \bibnamefont {Batista}}, \bibinfo {author}
  {\bibfnamefont {A.}~\bibnamefont {del Campo}}, \bibinfo {author}
  {\bibfnamefont {E.~J.}\ \bibnamefont {Torres-Herrera}},\ and\ \bibinfo
  {author} {\bibfnamefont {L.~F.}\ \bibnamefont {Santos}},\ }\href@noop {}
  {\bibinfo {title} {Proposal for many-body quantum chaos detection}} (\bibinfo
  {year} {2024}{\natexlab{a}}),\ \Eprint {https://arxiv.org/abs/2401.01401}
  {arXiv:2401.01401 [cond-mat.stat-mech]} \BibitemShut {NoStop}%
\bibitem [{\citenamefont {Lerma-Hern\'andez}\ \emph {et~al.}(2019)\citenamefont
  {Lerma-Hern\'andez}, \citenamefont {Villase\~nor}, \citenamefont
  {Bastarrachea-Magnani}, \citenamefont {Torres-Herrera}, \citenamefont
  {Santos},\ and\ \citenamefont {Hirsch}}]{Lerma2019}%
  \BibitemOpen
  \bibfield  {author} {\bibinfo {author} {\bibfnamefont {S.}~\bibnamefont
  {Lerma-Hern\'andez}}, \bibinfo {author} {\bibfnamefont {D.}~\bibnamefont
  {Villase\~nor}}, \bibinfo {author} {\bibfnamefont {M.~A.}\ \bibnamefont
  {Bastarrachea-Magnani}}, \bibinfo {author} {\bibfnamefont {E.~J.}\
  \bibnamefont {Torres-Herrera}}, \bibinfo {author} {\bibfnamefont {L.~F.}\
  \bibnamefont {Santos}},\ and\ \bibinfo {author} {\bibfnamefont {J.~G.}\
  \bibnamefont {Hirsch}},\ }\bibfield  {title} {\bibinfo {title} {Dynamical
  signatures of quantum chaos and relaxation time scales in a spin-boson
  system},\ }\href {https://doi.org/10.1103/PhysRevE.100.012218} {\bibfield
  {journal} {\bibinfo  {journal} {Phys. Rev. E}\ }\textbf {\bibinfo {volume}
  {100}},\ \bibinfo {pages} {012218} (\bibinfo {year} {2019})}\BibitemShut
  {NoStop}%
\bibitem [{\citenamefont {Santos}\ \emph {et~al.}(2020)\citenamefont {Santos},
  \citenamefont {P\'erez-Bernal},\ and\ \citenamefont
  {Torres-Herrera}}]{Santos2020}%
  \BibitemOpen
  \bibfield  {author} {\bibinfo {author} {\bibfnamefont {L.~F.}\ \bibnamefont
  {Santos}}, \bibinfo {author} {\bibfnamefont {F.}~\bibnamefont
  {P\'erez-Bernal}},\ and\ \bibinfo {author} {\bibfnamefont {E.~J.}\
  \bibnamefont {Torres-Herrera}},\ }\bibfield  {title} {\bibinfo {title} {Speck
  of chaos},\ }\href {https://doi.org/10.1103/PhysRevResearch.2.043034}
  {\bibfield  {journal} {\bibinfo  {journal} {Phys. Rev. Res.}\ }\textbf
  {\bibinfo {volume} {2}},\ \bibinfo {pages} {043034} (\bibinfo {year}
  {2020})}\BibitemShut {NoStop}%
\bibitem [{\citenamefont {del Campo}\ \emph {et~al.}(2017)\citenamefont {del
  Campo}, \citenamefont {Molina-Vilaplana},\ and\ \citenamefont
  {Sonner}}]{Campo2018a}%
  \BibitemOpen
  \bibfield  {author} {\bibinfo {author} {\bibfnamefont {A.}~\bibnamefont {del
  Campo}}, \bibinfo {author} {\bibfnamefont {J.}~\bibnamefont
  {Molina-Vilaplana}},\ and\ \bibinfo {author} {\bibfnamefont {J.}~\bibnamefont
  {Sonner}},\ }\bibfield  {title} {\bibinfo {title} {Scrambling the spectral
  form factor: Unitarity constraints and exact results},\ }\href
  {https://doi.org/10.1103/PhysRevD.95.126008} {\bibfield  {journal} {\bibinfo
  {journal} {Phys. Rev. D}\ }\textbf {\bibinfo {volume} {95}},\ \bibinfo
  {pages} {126008} (\bibinfo {year} {2017})}\BibitemShut {NoStop}%
\bibitem [{\citenamefont {Lezama}\ \emph {et~al.}(2021)\citenamefont {Lezama},
  \citenamefont {Torres-Herrera}, \citenamefont {P\'erez-Bernal}, \citenamefont
  {Bar~Lev},\ and\ \citenamefont {Santos}}]{Lezama2021}%
  \BibitemOpen
  \bibfield  {author} {\bibinfo {author} {\bibfnamefont {T.~L.~M.}\
  \bibnamefont {Lezama}}, \bibinfo {author} {\bibfnamefont {E.~J.}\
  \bibnamefont {Torres-Herrera}}, \bibinfo {author} {\bibfnamefont
  {F.}~\bibnamefont {P\'erez-Bernal}}, \bibinfo {author} {\bibfnamefont
  {Y.}~\bibnamefont {Bar~Lev}},\ and\ \bibinfo {author} {\bibfnamefont {L.~F.}\
  \bibnamefont {Santos}},\ }\bibfield  {title} {\bibinfo {title} {Equilibration
  time in many-body quantum systems},\ }\href
  {https://doi.org/10.1103/PhysRevB.104.085117} {\bibfield  {journal} {\bibinfo
   {journal} {Phys. Rev. B}\ }\textbf {\bibinfo {volume} {104}},\ \bibinfo
  {pages} {085117} (\bibinfo {year} {2021})}\BibitemShut {NoStop}%
\bibitem [{\citenamefont {Das}\ and\ \citenamefont
  {Ghosh}(2022{\natexlab{a}})}]{Das2022a}%
  \BibitemOpen
  \bibfield  {author} {\bibinfo {author} {\bibfnamefont {A.~K.}\ \bibnamefont
  {Das}}\ and\ \bibinfo {author} {\bibfnamefont {A.}~\bibnamefont {Ghosh}},\
  }\bibfield  {title} {\bibinfo {title} {Chaos due to symmetry-breaking in
  deformed {P}oisson ensemble},\ }\href
  {https://iopscience.iop.org/article/10.1088/1742-5468/ac70dd} {\bibfield
  {journal} {\bibinfo  {journal} {J. Stat. Mech.}\ }\textbf {\bibinfo {volume}
  {2022}},\ \bibinfo {pages} {063101} (\bibinfo {year}
  {2022}{\natexlab{a}})}\BibitemShut {NoStop}%
\bibitem [{\citenamefont {Das}\ and\ \citenamefont
  {Ghosh}(2022{\natexlab{b}})}]{Das2022b}%
  \BibitemOpen
  \bibfield  {author} {\bibinfo {author} {\bibfnamefont {A.~K.}\ \bibnamefont
  {Das}}\ and\ \bibinfo {author} {\bibfnamefont {A.}~\bibnamefont {Ghosh}},\
  }\bibfield  {title} {\bibinfo {title} {Transport in deformed centrosymmetric
  networks},\ }\href {https://doi.org/10.1103/PhysRevE.106.064112} {\bibfield
  {journal} {\bibinfo  {journal} {Phys. Rev. E}\ }\textbf {\bibinfo {volume}
  {106}},\ \bibinfo {pages} {064112} (\bibinfo {year}
  {2022}{\natexlab{b}})}\BibitemShut {NoStop}%
\bibitem [{\citenamefont {Das}\ and\ \citenamefont {Ghosh}(2023)}]{Das2023b}%
  \BibitemOpen
  \bibfield  {author} {\bibinfo {author} {\bibfnamefont {A.~K.}\ \bibnamefont
  {Das}}\ and\ \bibinfo {author} {\bibfnamefont {A.}~\bibnamefont {Ghosh}},\
  }\bibfield  {title} {\bibinfo {title} {Dynamical signatures of chaos to
  integrability crossover in $2 \times 2$ generalized random matrix
  ensembles},\ }\bibfield  {journal} {\bibinfo  {journal} {J. Phys. A}\ }\href
  {https://doi.org/10.1088/1751-8121/ad0b5a} {10.1088/1751-8121/ad0b5a}
  (\bibinfo {year} {2023})\BibitemShut {NoStop}%
\bibitem [{\citenamefont {Shir}\ \emph {et~al.}(2024)\citenamefont {Shir},
  \citenamefont {Martinez-Azcona},\ and\ \citenamefont
  {Chenu}}]{shir2023range}%
  \BibitemOpen
  \bibfield  {author} {\bibinfo {author} {\bibfnamefont {R.}~\bibnamefont
  {Shir}}, \bibinfo {author} {\bibfnamefont {P.}~\bibnamefont
  {Martinez-Azcona}},\ and\ \bibinfo {author} {\bibfnamefont {A.}~\bibnamefont
  {Chenu}},\ }\href@noop {} {\bibinfo {title} {Full range spectral correlations
  and their spectral form factors in chaotic and integrable models}} (\bibinfo
  {year} {2024}),\ \Eprint {https://arxiv.org/abs/2311.09292} {arXiv:2311.09292
  [quant-ph]} \BibitemShut {NoStop}%
\bibitem [{\citenamefont {Cotler}\ \emph
  {et~al.}(2017{\natexlab{a}})\citenamefont {Cotler}, \citenamefont {Gur-Ari},
  \citenamefont {Hanada}, \citenamefont {Polchinski}, \citenamefont {Saad},
  \citenamefont {Shenker}, \citenamefont {Stanford}, \citenamefont
  {Streicher},\ and\ \citenamefont {Tezuka}}]{Cotler2017}%
  \BibitemOpen
  \bibfield  {author} {\bibinfo {author} {\bibfnamefont {J.~S.}\ \bibnamefont
  {Cotler}}, \bibinfo {author} {\bibfnamefont {G.}~\bibnamefont {Gur-Ari}},
  \bibinfo {author} {\bibfnamefont {M.}~\bibnamefont {Hanada}}, \bibinfo
  {author} {\bibfnamefont {J.}~\bibnamefont {Polchinski}}, \bibinfo {author}
  {\bibfnamefont {P.}~\bibnamefont {Saad}}, \bibinfo {author} {\bibfnamefont
  {S.~H.}\ \bibnamefont {Shenker}}, \bibinfo {author} {\bibfnamefont
  {D.}~\bibnamefont {Stanford}}, \bibinfo {author} {\bibfnamefont
  {A.}~\bibnamefont {Streicher}},\ and\ \bibinfo {author} {\bibfnamefont
  {M.}~\bibnamefont {Tezuka}},\ }\bibfield  {title} {\bibinfo {title} {Black
  holes and random matrices},\ }\href {https://doi.org/10.1007/JHEP05(2017)118}
  {\bibfield  {journal} {\bibinfo  {journal} {J. High Energy Phys.}\ }\textbf
  {\bibinfo {volume} {2017}}\bibinfo  {number} { (5)},\ \bibinfo {pages}
  {118}}\BibitemShut {NoStop}%
\bibitem [{\citenamefont {Cotler}\ \emph
  {et~al.}(2017{\natexlab{b}})\citenamefont {Cotler}, \citenamefont
  {Hunter-Jones}, \citenamefont {Liu},\ and\ \citenamefont
  {Yoshida}}]{Cotler2017GUE}%
  \BibitemOpen
\bibfield  {number} {  }\bibfield  {author} {\bibinfo {author} {\bibfnamefont
  {J.}~\bibnamefont {Cotler}}, \bibinfo {author} {\bibfnamefont
  {N.}~\bibnamefont {Hunter-Jones}}, \bibinfo {author} {\bibfnamefont
  {J.}~\bibnamefont {Liu}},\ and\ \bibinfo {author} {\bibfnamefont
  {B.}~\bibnamefont {Yoshida}},\ }\bibfield  {title} {\bibinfo {title} {Chaos,
  complexity, and random matrices},\ }\href
  {https://doi.org/10.1007/JHEP11(2017)048} {\bibfield  {journal} {\bibinfo
  {journal} {J. High Energy Phys.}\ }\textbf {\bibinfo {volume} {2017}}\bibinfo
   {number} { (11)},\ \bibinfo {pages} {48}}\BibitemShut {NoStop}%
\bibitem [{\citenamefont {\v{S}untajs}\ \emph {et~al.}(2020)\citenamefont
  {\v{S}untajs}, \citenamefont {Bon\v{c}a}, \citenamefont {Prosen},\ and\
  \citenamefont {Vidmar}}]{Suntajs2020}%
  \BibitemOpen
\bibfield  {number} {  }\bibfield  {author} {\bibinfo {author} {\bibfnamefont
  {J.}~\bibnamefont {\v{S}untajs}}, \bibinfo {author} {\bibfnamefont
  {J.}~\bibnamefont {Bon\v{c}a}}, \bibinfo {author} {\bibfnamefont
  {T.}~\bibnamefont {Prosen}},\ and\ \bibinfo {author} {\bibfnamefont
  {L.}~\bibnamefont {Vidmar}},\ }\bibfield  {title} {\bibinfo {title} {Quantum
  chaos challenges many-body localization},\ }\href
  {https://doi.org/10.1103/PhysRevE.102.062144} {\bibfield  {journal} {\bibinfo
   {journal} {Phys. Rev. E}\ }\textbf {\bibinfo {volume} {102}},\ \bibinfo
  {pages} {062144} (\bibinfo {year} {2020})}\BibitemShut {NoStop}%
\bibitem [{\citenamefont {Dag}\ \emph {et~al.}(2023)\citenamefont {Dag},
  \citenamefont {Mistakidis}, \citenamefont {Chan},\ and\ \citenamefont
  {Sadeghpour}}]{Dag2023}%
  \BibitemOpen
  \bibfield  {author} {\bibinfo {author} {\bibfnamefont {C.~B.}\ \bibnamefont
  {Dag}}, \bibinfo {author} {\bibfnamefont {S.~I.}\ \bibnamefont {Mistakidis}},
  \bibinfo {author} {\bibfnamefont {A.}~\bibnamefont {Chan}},\ and\ \bibinfo
  {author} {\bibfnamefont {H.~R.}\ \bibnamefont {Sadeghpour}},\ }\bibfield
  {title} {\bibinfo {title} {Many-body quantum chaos in stroboscopically-driven
  cold atoms},\ }\href {https://doi.org/10.1038/s42005-023-01258-1} {\bibfield
  {journal} {\bibinfo  {journal} {Comm. Phys.}\ }\textbf {\bibinfo {volume}
  {6}},\ \bibinfo {pages} {136} (\bibinfo {year} {2023})}\BibitemShut {NoStop}%
\bibitem [{\citenamefont {Dong}\ \emph {et~al.}(2024)\citenamefont {Dong},
  \citenamefont {Zhang}, \citenamefont {Dag}, \citenamefont {Gao},
  \citenamefont {Wang}, \citenamefont {Deng}, \citenamefont {Zhang},
  \citenamefont {Chen}, \citenamefont {Xu}, \citenamefont {Wang}, \citenamefont
  {Wu}, \citenamefont {Zhang}, \citenamefont {Jin}, \citenamefont {Zhu},
  \citenamefont {Zhang}, \citenamefont {Zou}, \citenamefont {Tan},
  \citenamefont {Cui}, \citenamefont {Zhu}, \citenamefont {Shen}, \citenamefont
  {Li}, \citenamefont {Zhong}, \citenamefont {Bao}, \citenamefont {Li},
  \citenamefont {Wang}, \citenamefont {Guo}, \citenamefont {Song},
  \citenamefont {Liu}, \citenamefont {Chan}, \citenamefont {Ying},\ and\
  \citenamefont {Wang}}]{dong2024measuring}%
  \BibitemOpen
  \bibfield  {author} {\bibinfo {author} {\bibfnamefont {H.}~\bibnamefont
  {Dong}}, \bibinfo {author} {\bibfnamefont {P.}~\bibnamefont {Zhang}},
  \bibinfo {author} {\bibfnamefont {C.~B.}\ \bibnamefont {Dag}}, \bibinfo
  {author} {\bibfnamefont {Y.}~\bibnamefont {Gao}}, \bibinfo {author}
  {\bibfnamefont {N.}~\bibnamefont {Wang}}, \bibinfo {author} {\bibfnamefont
  {J.}~\bibnamefont {Deng}}, \bibinfo {author} {\bibfnamefont {X.}~\bibnamefont
  {Zhang}}, \bibinfo {author} {\bibfnamefont {J.}~\bibnamefont {Chen}},
  \bibinfo {author} {\bibfnamefont {S.}~\bibnamefont {Xu}}, \bibinfo {author}
  {\bibfnamefont {K.}~\bibnamefont {Wang}}, \bibinfo {author} {\bibfnamefont
  {Y.}~\bibnamefont {Wu}}, \bibinfo {author} {\bibfnamefont {C.}~\bibnamefont
  {Zhang}}, \bibinfo {author} {\bibfnamefont {F.}~\bibnamefont {Jin}}, \bibinfo
  {author} {\bibfnamefont {X.}~\bibnamefont {Zhu}}, \bibinfo {author}
  {\bibfnamefont {A.}~\bibnamefont {Zhang}}, \bibinfo {author} {\bibfnamefont
  {Y.}~\bibnamefont {Zou}}, \bibinfo {author} {\bibfnamefont {Z.}~\bibnamefont
  {Tan}}, \bibinfo {author} {\bibfnamefont {Z.}~\bibnamefont {Cui}}, \bibinfo
  {author} {\bibfnamefont {Z.}~\bibnamefont {Zhu}}, \bibinfo {author}
  {\bibfnamefont {F.}~\bibnamefont {Shen}}, \bibinfo {author} {\bibfnamefont
  {T.}~\bibnamefont {Li}}, \bibinfo {author} {\bibfnamefont {J.}~\bibnamefont
  {Zhong}}, \bibinfo {author} {\bibfnamefont {Z.}~\bibnamefont {Bao}}, \bibinfo
  {author} {\bibfnamefont {H.}~\bibnamefont {Li}}, \bibinfo {author}
  {\bibfnamefont {Z.}~\bibnamefont {Wang}}, \bibinfo {author} {\bibfnamefont
  {Q.}~\bibnamefont {Guo}}, \bibinfo {author} {\bibfnamefont {C.}~\bibnamefont
  {Song}}, \bibinfo {author} {\bibfnamefont {F.}~\bibnamefont {Liu}}, \bibinfo
  {author} {\bibfnamefont {A.}~\bibnamefont {Chan}}, \bibinfo {author}
  {\bibfnamefont {L.}~\bibnamefont {Ying}},\ and\ \bibinfo {author}
  {\bibfnamefont {H.}~\bibnamefont {Wang}},\ }\href@noop {} {\bibinfo {title}
  {Measuring spectral form factor in many-body chaotic and localized phases of
  quantum processors}} (\bibinfo {year} {2024}),\ \Eprint
  {https://arxiv.org/abs/2403.16935} {arXiv:2403.16935} \BibitemShut {NoStop}%
\bibitem [{\citenamefont {del Campo}\ \emph {et~al.}(2018)\citenamefont {del
  Campo}, \citenamefont {Molina-Vilaplana}, \citenamefont {Santos},\ and\
  \citenamefont {Sonner}}]{Campo2018}%
  \BibitemOpen
  \bibfield  {author} {\bibinfo {author} {\bibfnamefont {A.}~\bibnamefont {del
  Campo}}, \bibinfo {author} {\bibfnamefont {J.}~\bibnamefont
  {Molina-Vilaplana}}, \bibinfo {author} {\bibfnamefont {L.}~\bibnamefont
  {Santos}},\ and\ \bibinfo {author} {\bibfnamefont {J.}~\bibnamefont
  {Sonner}},\ }\bibfield  {title} {\bibinfo {title} {Decay of a
  thermofield-double state in chaotic quantum systems},\ }\href
  {https://doi.org/10.1140/epjst/e2018-00083-5} {\bibfield  {journal} {\bibinfo
   {journal} {Eur. Phys. J. Spec. Top.}\ }\textbf {\bibinfo {volume} {227}},\
  \bibinfo {pages} {247} (\bibinfo {year} {2018})}\BibitemShut {NoStop}%
\bibitem [{\citenamefont {Jozsa}(1994)}]{Jozsa1994}%
  \BibitemOpen
  \bibfield  {author} {\bibinfo {author} {\bibfnamefont {R.}~\bibnamefont
  {Jozsa}},\ }\bibfield  {title} {\bibinfo {title} {Fidelity for mixed quantum
  states},\ }\href {https://doi.org/10.1080/09500349414552171} {\bibfield
  {journal} {\bibinfo  {journal} {J. Mod. Opt.}\ }\textbf {\bibinfo {volume}
  {41}},\ \bibinfo {pages} {2315} (\bibinfo {year} {1994})}\BibitemShut
  {NoStop}%
\bibitem [{\citenamefont {Carrera-N{\'u}{\~n}ez}\ \emph
  {et~al.}(2022)\citenamefont {Carrera-N{\'u}{\~n}ez}, \citenamefont
  {Mart{\'\i}nez-Arg{\"u}ello}, \citenamefont {Torres},\ and\ \citenamefont
  {Torres-Herrera}}]{Carrera2022}%
  \BibitemOpen
  \bibfield  {author} {\bibinfo {author} {\bibfnamefont {M.}~\bibnamefont
  {Carrera-N{\'u}{\~n}ez}}, \bibinfo {author} {\bibfnamefont {A.}~\bibnamefont
  {Mart{\'\i}nez-Arg{\"u}ello}}, \bibinfo {author} {\bibfnamefont
  {J.}~\bibnamefont {Torres}},\ and\ \bibinfo {author} {\bibfnamefont
  {E.}~\bibnamefont {Torres-Herrera}},\ }\bibfield  {title} {\bibinfo {title}
  {Onset of universality in the dynamical mixing of a pure state},\ }\href
  {https://doi.org/10.1088/1751-8121/ac9f8b} {\bibfield  {journal} {\bibinfo
  {journal} {J. Phys. A}\ }\textbf {\bibinfo {volume} {55}},\ \bibinfo {pages}
  {455303} (\bibinfo {year} {2022})}\BibitemShut {NoStop}%
\bibitem [{\citenamefont {T\'avora}\ \emph {et~al.}(2016)\citenamefont
  {T\'avora}, \citenamefont {Torres-Herrera},\ and\ \citenamefont
  {Santos}}]{Tavora2016}%
  \BibitemOpen
  \bibfield  {author} {\bibinfo {author} {\bibfnamefont {M.}~\bibnamefont
  {T\'avora}}, \bibinfo {author} {\bibfnamefont {E.~J.}\ \bibnamefont
  {Torres-Herrera}},\ and\ \bibinfo {author} {\bibfnamefont {L.~F.}\
  \bibnamefont {Santos}},\ }\bibfield  {title} {\bibinfo {title} {Inevitable
  power-law behavior of isolated many-body quantum systems and how it
  anticipates thermalization},\ }\href
  {https://doi.org/10.1103/PhysRevA.94.041603} {\bibfield  {journal} {\bibinfo
  {journal} {Phys. Rev. A}\ }\textbf {\bibinfo {volume} {94}},\ \bibinfo
  {pages} {041603} (\bibinfo {year} {2016})}\BibitemShut {NoStop}%
\bibitem [{\citenamefont {T\'avora}\ \emph {et~al.}(2017)\citenamefont
  {T\'avora}, \citenamefont {Torres-Herrera},\ and\ \citenamefont
  {Santos}}]{Tavora2017}%
  \BibitemOpen
  \bibfield  {author} {\bibinfo {author} {\bibfnamefont {M.}~\bibnamefont
  {T\'avora}}, \bibinfo {author} {\bibfnamefont {E.~J.}\ \bibnamefont
  {Torres-Herrera}},\ and\ \bibinfo {author} {\bibfnamefont {L.~F.}\
  \bibnamefont {Santos}},\ }\bibfield  {title} {\bibinfo {title} {Power-law
  decay exponents: A dynamical criterion for predicting thermalization},\
  }\href {https://doi.org/10.1103/PhysRevA.95.013604} {\bibfield  {journal}
  {\bibinfo  {journal} {Phys. Rev. A}\ }\textbf {\bibinfo {volume} {95}},\
  \bibinfo {pages} {013604} (\bibinfo {year} {2017})}\BibitemShut {NoStop}%
\bibitem [{\citenamefont {Mirlin}\ \emph {et~al.}(1996)\citenamefont {Mirlin},
  \citenamefont {Fyodorov}, \citenamefont {Dittes}, \citenamefont {Quezada},\
  and\ \citenamefont {Seligman}}]{PhysRevE.54.3221}%
  \BibitemOpen
  \bibfield  {author} {\bibinfo {author} {\bibfnamefont {A.~D.}\ \bibnamefont
  {Mirlin}}, \bibinfo {author} {\bibfnamefont {Y.~V.}\ \bibnamefont
  {Fyodorov}}, \bibinfo {author} {\bibfnamefont {F.-M.}\ \bibnamefont
  {Dittes}}, \bibinfo {author} {\bibfnamefont {J.}~\bibnamefont {Quezada}},\
  and\ \bibinfo {author} {\bibfnamefont {T.~H.}\ \bibnamefont {Seligman}},\
  }\bibfield  {title} {\bibinfo {title} {Transition from localized to extended
  eigenstates in the ensemble of power-law random banded matrices},\ }\href
  {https://doi.org/10.1103/PhysRevE.54.3221} {\bibfield  {journal} {\bibinfo
  {journal} {Phys. Rev. E}\ }\textbf {\bibinfo {volume} {54}},\ \bibinfo
  {pages} {3221} (\bibinfo {year} {1996})}\BibitemShut {NoStop}%
\bibitem [{\citenamefont {Varga}\ and\ \citenamefont
  {Braun}(2000)}]{PhysRevB.61.R11859}%
  \BibitemOpen
  \bibfield  {author} {\bibinfo {author} {\bibfnamefont {I.}~\bibnamefont
  {Varga}}\ and\ \bibinfo {author} {\bibfnamefont {D.}~\bibnamefont {Braun}},\
  }\bibfield  {title} {\bibinfo {title} {Critical statistics in a power-law
  random-banded matrix ensemble},\ }\href
  {https://doi.org/10.1103/PhysRevB.61.R11859} {\bibfield  {journal} {\bibinfo
  {journal} {Phys. Rev. B}\ }\textbf {\bibinfo {volume} {61}},\ \bibinfo
  {pages} {R11859} (\bibinfo {year} {2000})}\BibitemShut {NoStop}%
\bibitem [{\citenamefont {Rosenzweig}\ and\ \citenamefont
  {Porter}(1960)}]{Rosenzweig1960}%
  \BibitemOpen
  \bibfield  {author} {\bibinfo {author} {\bibfnamefont {N.}~\bibnamefont
  {Rosenzweig}}\ and\ \bibinfo {author} {\bibfnamefont {C.~E.}\ \bibnamefont
  {Porter}},\ }\bibfield  {title} {\bibinfo {title} {Repulsion of energy levels
  in complex atomic spectra},\ }\href
  {https://doi.org/10.1103/PhysRev.120.1698} {\bibfield  {journal} {\bibinfo
  {journal} {Phys. Rev.}\ }\textbf {\bibinfo {volume} {120}},\ \bibinfo {pages}
  {1698} (\bibinfo {year} {1960})}\BibitemShut {NoStop}%
\bibitem [{\citenamefont {de~Carvalho}\ \emph {et~al.}(2007)\citenamefont
  {de~Carvalho}, \citenamefont {Hussein}, \citenamefont {Pato},\ and\
  \citenamefont {Sargeant}}]{Carvalho2007}%
  \BibitemOpen
  \bibfield  {author} {\bibinfo {author} {\bibfnamefont {J.~X.}\ \bibnamefont
  {de~Carvalho}}, \bibinfo {author} {\bibfnamefont {M.~S.}\ \bibnamefont
  {Hussein}}, \bibinfo {author} {\bibfnamefont {M.~P.}\ \bibnamefont {Pato}},\
  and\ \bibinfo {author} {\bibfnamefont {A.~J.}\ \bibnamefont {Sargeant}},\
  }\bibfield  {title} {\bibinfo {title} {Symmetry-breaking study with deformed
  ensembles},\ }\href {https://doi.org/10.1103/PhysRevE.76.066212} {\bibfield
  {journal} {\bibinfo  {journal} {Phys. Rev. E}\ }\textbf {\bibinfo {volume}
  {76}},\ \bibinfo {pages} {066212} (\bibinfo {year} {2007})}\BibitemShut
  {NoStop}%
\bibitem [{\citenamefont {Kravtsov}\ \emph {et~al.}(2015)\citenamefont
  {Kravtsov}, \citenamefont {Khaymovich}, \citenamefont {Cuevas},\ and\
  \citenamefont {Amini}}]{Kravtsov2015}%
  \BibitemOpen
  \bibfield  {author} {\bibinfo {author} {\bibfnamefont {V.~E.}\ \bibnamefont
  {Kravtsov}}, \bibinfo {author} {\bibfnamefont {I.~M.}\ \bibnamefont
  {Khaymovich}}, \bibinfo {author} {\bibfnamefont {E.}~\bibnamefont {Cuevas}},\
  and\ \bibinfo {author} {\bibfnamefont {M.}~\bibnamefont {Amini}},\ }\bibfield
   {title} {\bibinfo {title} {A random matrix model with localization and
  ergodic transitions},\ }\href
  {https://doi.org/10.1088/1367-2630/17/12/122002} {\bibfield  {journal}
  {\bibinfo  {journal} {New J. Phys.}\ }\textbf {\bibinfo {volume} {17}},\
  \bibinfo {pages} {122002} (\bibinfo {year} {2015})}\BibitemShut {NoStop}%
\bibitem [{\citenamefont {Monthus}(2017)}]{Monthus2017}%
  \BibitemOpen
  \bibfield  {author} {\bibinfo {author} {\bibfnamefont {C.}~\bibnamefont
  {Monthus}},\ }\bibfield  {title} {\bibinfo {title} {Multifractality of
  eigenstates in the delocalized non-ergodic phase of some random matrix
  models: {W}igner-{W}eisskopf approach},\ }\href
  {https://doi.org/10.1088/1751-8121/aa77e1} {\bibfield  {journal} {\bibinfo
  {journal} {J. Phys. A}\ }\textbf {\bibinfo {volume} {50}},\ \bibinfo {pages}
  {295101} (\bibinfo {year} {2017})}\BibitemShut {NoStop}%
\bibitem [{\citenamefont {Pino}\ \emph {et~al.}(2019)\citenamefont {Pino},
  \citenamefont {Tabanera},\ and\ \citenamefont {Serna}}]{Pino2019}%
  \BibitemOpen
  \bibfield  {author} {\bibinfo {author} {\bibfnamefont {M.}~\bibnamefont
  {Pino}}, \bibinfo {author} {\bibfnamefont {J.}~\bibnamefont {Tabanera}},\
  and\ \bibinfo {author} {\bibfnamefont {P.}~\bibnamefont {Serna}},\ }\bibfield
   {title} {\bibinfo {title} {From ergodic to non-ergodic chaos in
  {R}osenzweig-{P}orter model},\ }\href
  {https://doi.org/10.1088/1751-8121/ab4b76} {\bibfield  {journal} {\bibinfo
  {journal} {J. Phys. A}\ }\textbf {\bibinfo {volume} {52}},\ \bibinfo {pages}
  {475101} (\bibinfo {year} {2019})}\BibitemShut {NoStop}%
\bibitem [{\citenamefont {von Soosten}\ and\ \citenamefont
  {Warzel}(2019)}]{Soosten2019}%
  \BibitemOpen
  \bibfield  {author} {\bibinfo {author} {\bibfnamefont {P.}~\bibnamefont {von
  Soosten}}\ and\ \bibinfo {author} {\bibfnamefont {S.}~\bibnamefont
  {Warzel}},\ }\bibfield  {title} {\bibinfo {title} {Non-ergodic delocalization
  in the {R}osenzweig-{P}orter model},\ }\href
  {https://doi.org/10.1007/s11005-018-1131-} {\bibfield  {journal} {\bibinfo
  {journal} {Lett. Math. Phys.}\ }\textbf {\bibinfo {volume} {109}},\ \bibinfo
  {pages} {905} (\bibinfo {year} {2019})}\BibitemShut {NoStop}%
\bibitem [{\citenamefont {Zhang}\ \emph {et~al.}(2023)\citenamefont {Zhang},
  \citenamefont {Zhang}, \citenamefont {Che},\ and\ \citenamefont
  {Dietz}}]{Zhang2023}%
  \BibitemOpen
  \bibfield  {author} {\bibinfo {author} {\bibfnamefont {X.}~\bibnamefont
  {Zhang}}, \bibinfo {author} {\bibfnamefont {W.}~\bibnamefont {Zhang}},
  \bibinfo {author} {\bibfnamefont {J.}~\bibnamefont {Che}},\ and\ \bibinfo
  {author} {\bibfnamefont {B.}~\bibnamefont {Dietz}},\ }\bibfield  {title}
  {\bibinfo {title} {Experimental test of the {R}osenzweig-{P}orter model for
  the transition from {P}oisson to {G}aussian unitary ensemble statistics},\
  }\href {https://doi.org/10.1103/PhysRevE.108.044211} {\bibfield  {journal}
  {\bibinfo  {journal} {Phys. Rev. E}\ }\textbf {\bibinfo {volume} {108}},\
  \bibinfo {pages} {044211} (\bibinfo {year} {2023})}\BibitemShut {NoStop}%
\bibitem [{\citenamefont {Nosov}\ \emph {et~al.}(2019)\citenamefont {Nosov},
  \citenamefont {Khaymovich},\ and\ \citenamefont {Kravtsov}}]{Nosov2019}%
  \BibitemOpen
  \bibfield  {author} {\bibinfo {author} {\bibfnamefont {P.~A.}\ \bibnamefont
  {Nosov}}, \bibinfo {author} {\bibfnamefont {I.~M.}\ \bibnamefont
  {Khaymovich}},\ and\ \bibinfo {author} {\bibfnamefont {V.~E.}\ \bibnamefont
  {Kravtsov}},\ }\bibfield  {title} {\bibinfo {title} {Correlation-induced
  localization},\ }\href {https://doi.org/10.1103/PhysRevB.99.104203}
  {\bibfield  {journal} {\bibinfo  {journal} {Phys. Rev. B}\ }\textbf {\bibinfo
  {volume} {99}},\ \bibinfo {pages} {104203} (\bibinfo {year}
  {2019})}\BibitemShut {NoStop}%
\bibitem [{\citenamefont {Cizeau}\ and\ \citenamefont
  {Bouchaud}(1994)}]{Cizeau1994}%
  \BibitemOpen
  \bibfield  {author} {\bibinfo {author} {\bibfnamefont {P.}~\bibnamefont
  {Cizeau}}\ and\ \bibinfo {author} {\bibfnamefont {J.~P.}\ \bibnamefont
  {Bouchaud}},\ }\bibfield  {title} {\bibinfo {title} {Theory of {L}\'evy
  matrices},\ }\href {https://doi.org/10.1103/PhysRevE.50.1810} {\bibfield
  {journal} {\bibinfo  {journal} {Phys. Rev. E}\ }\textbf {\bibinfo {volume}
  {50}},\ \bibinfo {pages} {1810} (\bibinfo {year} {1994})}\BibitemShut
  {NoStop}%
\bibitem [{\citenamefont {Kutlin}\ and\ \citenamefont
  {Khaymovich}(2021)}]{Kutlin2021}%
  \BibitemOpen
  \bibfield  {author} {\bibinfo {author} {\bibfnamefont {A.~G.}\ \bibnamefont
  {Kutlin}}\ and\ \bibinfo {author} {\bibfnamefont {I.~M.}\ \bibnamefont
  {Khaymovich}},\ }\bibfield  {title} {\bibinfo {title} {{Emergent fractal
  phase in energy stratified random models}},\ }\href
  {https://doi.org/10.21468/SciPostPhys.11.6.101} {\bibfield  {journal}
  {\bibinfo  {journal} {SciPost Phys.}\ }\textbf {\bibinfo {volume} {11}},\
  \bibinfo {pages} {101} (\bibinfo {year} {2021})}\BibitemShut {NoStop}%
\bibitem [{\citenamefont {Tang}\ and\ \citenamefont
  {Khaymovich}(2022)}]{Tang2022}%
  \BibitemOpen
  \bibfield  {author} {\bibinfo {author} {\bibfnamefont {W.}~\bibnamefont
  {Tang}}\ and\ \bibinfo {author} {\bibfnamefont {I.~M.}\ \bibnamefont
  {Khaymovich}},\ }\bibfield  {title} {\bibinfo {title} {Non-ergodic
  delocalized phase with {P}oisson level statistics},\ }\href
  {https://doi.org/10.22331/q-2022-06-09-733} {\bibfield  {journal} {\bibinfo
  {journal} {Quantum}\ }\textbf {\bibinfo {volume} {6}},\ \bibinfo {pages}
  {733} (\bibinfo {year} {2022})}\BibitemShut {NoStop}%
\bibitem [{\citenamefont {Motamarri}\ \emph {et~al.}(2022)\citenamefont
  {Motamarri}, \citenamefont {Gorsky},\ and\ \citenamefont
  {Khaymovich}}]{Motamarri2021}%
  \BibitemOpen
  \bibfield  {author} {\bibinfo {author} {\bibfnamefont {V.~R.}\ \bibnamefont
  {Motamarri}}, \bibinfo {author} {\bibfnamefont {A.~S.}\ \bibnamefont
  {Gorsky}},\ and\ \bibinfo {author} {\bibfnamefont {I.~M.}\ \bibnamefont
  {Khaymovich}},\ }\bibfield  {title} {\bibinfo {title} {{Localization and
  fractality in disordered Russian Doll model}},\ }\href
  {https://doi.org/10.21468/SciPostPhys.13.5.117} {\bibfield  {journal}
  {\bibinfo  {journal} {SciPost Phys.}\ }\textbf {\bibinfo {volume} {13}},\
  \bibinfo {pages} {117} (\bibinfo {year} {2022})}\BibitemShut {NoStop}%
\bibitem [{\citenamefont {Das}\ and\ \citenamefont
  {Ghosh}(2022{\natexlab{c}})}]{Das2022c}%
  \BibitemOpen
  \bibfield  {author} {\bibinfo {author} {\bibfnamefont {A.~K.}\ \bibnamefont
  {Das}}\ and\ \bibinfo {author} {\bibfnamefont {A.}~\bibnamefont {Ghosh}},\
  }\bibfield  {title} {\bibinfo {title} {Nonergodic extended states in the
  $\beta$ ensemble},\ }\href {https://doi.org/10.1103/PhysRevE.105.054121}
  {\bibfield  {journal} {\bibinfo  {journal} {Phys. Rev. E}\ }\textbf {\bibinfo
  {volume} {105}},\ \bibinfo {pages} {054121} (\bibinfo {year}
  {2022}{\natexlab{c}})}\BibitemShut {NoStop}%
\bibitem [{\citenamefont {Das}\ \emph {et~al.}(2023)\citenamefont {Das},
  \citenamefont {Ghosh},\ and\ \citenamefont {Khaymovich}}]{Das2023a}%
  \BibitemOpen
  \bibfield  {author} {\bibinfo {author} {\bibfnamefont {A.~K.}\ \bibnamefont
  {Das}}, \bibinfo {author} {\bibfnamefont {A.}~\bibnamefont {Ghosh}},\ and\
  \bibinfo {author} {\bibfnamefont {I.~M.}\ \bibnamefont {Khaymovich}},\
  }\bibfield  {title} {\bibinfo {title} {Absence of mobility edge in
  short-range uncorrelated disordered model: Coexistence of localized and
  extended states},\ }\href {https://doi.org/10.1103/PhysRevLett.131.166401}
  {\bibfield  {journal} {\bibinfo  {journal} {Phys. Rev. Lett.}\ }\textbf
  {\bibinfo {volume} {131}},\ \bibinfo {pages} {166401} (\bibinfo {year}
  {2023})}\BibitemShut {NoStop}%
\bibitem [{\citenamefont {Das}\ \emph {et~al.}(2024{\natexlab{b}})\citenamefont
  {Das}, \citenamefont {Ghosh},\ and\ \citenamefont {Khaymovich}}]{Das2024a}%
  \BibitemOpen
  \bibfield  {author} {\bibinfo {author} {\bibfnamefont {A.~K.}\ \bibnamefont
  {Das}}, \bibinfo {author} {\bibfnamefont {A.}~\bibnamefont {Ghosh}},\ and\
  \bibinfo {author} {\bibfnamefont {I.~M.}\ \bibnamefont {Khaymovich}},\
  }\bibfield  {title} {\bibinfo {title} {Robust nonergodicity of the ground
  states in the $\ensuremath{\beta}$ ensemble},\ }\href
  {https://doi.org/10.1103/PhysRevB.109.064206} {\bibfield  {journal} {\bibinfo
   {journal} {Phys. Rev. B}\ }\textbf {\bibinfo {volume} {109}},\ \bibinfo
  {pages} {064206} (\bibinfo {year} {2024}{\natexlab{b}})}\BibitemShut
  {NoStop}%
\bibitem [{\citenamefont {Tikhonov}\ and\ \citenamefont
  {Mirlin}(2016)}]{Tikhonov2016}%
  \BibitemOpen
  \bibfield  {author} {\bibinfo {author} {\bibfnamefont {K.~S.}\ \bibnamefont
  {Tikhonov}}\ and\ \bibinfo {author} {\bibfnamefont {A.~D.}\ \bibnamefont
  {Mirlin}},\ }\bibfield  {title} {\bibinfo {title} {Fractality of wave
  functions on a cayley tree: Difference between tree and locally treelike
  graph without boundary},\ }\href {https://doi.org/10.1103/PhysRevB.94.184203}
  {\bibfield  {journal} {\bibinfo  {journal} {Phys. Rev. B}\ }\textbf {\bibinfo
  {volume} {94}},\ \bibinfo {pages} {184203} (\bibinfo {year}
  {2016})}\BibitemShut {NoStop}%
\bibitem [{\citenamefont {Luitz}\ \emph {et~al.}(2015)\citenamefont {Luitz},
  \citenamefont {Laflorencie},\ and\ \citenamefont {Alet}}]{Luitz2015}%
  \BibitemOpen
  \bibfield  {author} {\bibinfo {author} {\bibfnamefont {D.~J.}\ \bibnamefont
  {Luitz}}, \bibinfo {author} {\bibfnamefont {N.}~\bibnamefont {Laflorencie}},\
  and\ \bibinfo {author} {\bibfnamefont {F.}~\bibnamefont {Alet}},\ }\bibfield
  {title} {\bibinfo {title} {Many-body localization edge in the random-field
  {H}eisenberg chain},\ }\href {https://doi.org/10.1103/PhysRevB.91.081103}
  {\bibfield  {journal} {\bibinfo  {journal} {Phys. Rev. B}\ }\textbf {\bibinfo
  {volume} {91}},\ \bibinfo {pages} {081103} (\bibinfo {year}
  {2015})}\BibitemShut {NoStop}%
\bibitem [{\citenamefont {Luitz}\ \emph {et~al.}(2020)\citenamefont {Luitz},
  \citenamefont {Khaymovich},\ and\ \citenamefont {Lev}}]{Luitz2020}%
  \BibitemOpen
  \bibfield  {author} {\bibinfo {author} {\bibfnamefont {D.~J.}\ \bibnamefont
  {Luitz}}, \bibinfo {author} {\bibfnamefont {I.~M.}\ \bibnamefont
  {Khaymovich}},\ and\ \bibinfo {author} {\bibfnamefont {Y.~B.}\ \bibnamefont
  {Lev}},\ }\bibfield  {title} {\bibinfo {title} {{Multifractality and its role
  in anomalous transport in the disordered XXZ spin-chain}},\ }\href
  {https://doi.org/10.21468/SciPostPhysCore.2.2.006} {\bibfield  {journal}
  {\bibinfo  {journal} {SciPost Phys. Core}\ }\textbf {\bibinfo {volume} {2}},\
  \bibinfo {pages} {6} (\bibinfo {year} {2020})}\BibitemShut {NoStop}%
\bibitem [{\citenamefont {Pino}\ \emph {et~al.}(2017)\citenamefont {Pino},
  \citenamefont {Kravtsov}, \citenamefont {Altshuler},\ and\ \citenamefont
  {Ioffe}}]{Pino2017}%
  \BibitemOpen
  \bibfield  {author} {\bibinfo {author} {\bibfnamefont {M.}~\bibnamefont
  {Pino}}, \bibinfo {author} {\bibfnamefont {V.~E.}\ \bibnamefont {Kravtsov}},
  \bibinfo {author} {\bibfnamefont {B.~L.}\ \bibnamefont {Altshuler}},\ and\
  \bibinfo {author} {\bibfnamefont {L.~B.}\ \bibnamefont {Ioffe}},\ }\bibfield
  {title} {\bibinfo {title} {Multifractal metal in a disordered {J}osephson
  junctions array},\ }\href {https://doi.org/10.1103/PhysRevB.96.214205}
  {\bibfield  {journal} {\bibinfo  {journal} {Phys. Rev. B}\ }\textbf {\bibinfo
  {volume} {96}},\ \bibinfo {pages} {214205} (\bibinfo {year}
  {2017})}\BibitemShut {NoStop}%
\bibitem [{\citenamefont {Altshuler}\ \emph {et~al.}(1997)\citenamefont
  {Altshuler}, \citenamefont {Gefen}, \citenamefont {Kamenev},\ and\
  \citenamefont {Levitov}}]{Altshuler1997}%
  \BibitemOpen
  \bibfield  {author} {\bibinfo {author} {\bibfnamefont {B.~L.}\ \bibnamefont
  {Altshuler}}, \bibinfo {author} {\bibfnamefont {Y.}~\bibnamefont {Gefen}},
  \bibinfo {author} {\bibfnamefont {A.}~\bibnamefont {Kamenev}},\ and\ \bibinfo
  {author} {\bibfnamefont {L.~S.}\ \bibnamefont {Levitov}},\ }\bibfield
  {title} {\bibinfo {title} {Quasiparticle lifetime in a finite system: A
  nonperturbative approach},\ }\href
  {https://doi.org/10.1103/PhysRevLett.78.2803} {\bibfield  {journal} {\bibinfo
   {journal} {Phys. Rev. Lett.}\ }\textbf {\bibinfo {volume} {78}},\ \bibinfo
  {pages} {2803} (\bibinfo {year} {1997})}\BibitemShut {NoStop}%
\bibitem [{\citenamefont {Khaymovich}\ \emph {et~al.}(2020)\citenamefont
  {Khaymovich}, \citenamefont {Kravtsov}, \citenamefont {Altshuler},\ and\
  \citenamefont {Ioffe}}]{Khaymovich2020}%
  \BibitemOpen
  \bibfield  {author} {\bibinfo {author} {\bibfnamefont {I.~M.}\ \bibnamefont
  {Khaymovich}}, \bibinfo {author} {\bibfnamefont {V.~E.}\ \bibnamefont
  {Kravtsov}}, \bibinfo {author} {\bibfnamefont {B.~L.}\ \bibnamefont
  {Altshuler}},\ and\ \bibinfo {author} {\bibfnamefont {L.~B.}\ \bibnamefont
  {Ioffe}},\ }\bibfield  {title} {\bibinfo {title} {Fragile extended phases in
  the log-normal {R}osenzweig-{P}orter model},\ }\href
  {https://doi.org/10.1103/PhysRevResearch.2.043346} {\bibfield  {journal}
  {\bibinfo  {journal} {Phys. Rev. Res.}\ }\textbf {\bibinfo {volume} {2}},\
  \bibinfo {pages} {043346} (\bibinfo {year} {2020})}\BibitemShut {NoStop}%
\bibitem [{\citenamefont {Khaymovich}\ and\ \citenamefont
  {Kravtsov}(2021)}]{Khaymovich2021}%
  \BibitemOpen
  \bibfield  {author} {\bibinfo {author} {\bibfnamefont {I.~M.}\ \bibnamefont
  {Khaymovich}}\ and\ \bibinfo {author} {\bibfnamefont {V.~E.}\ \bibnamefont
  {Kravtsov}},\ }\bibfield  {title} {\bibinfo {title} {{Dynamical phases in a
  ``multifractal'' {R}osenzweig-{P}orter model}},\ }\href
  {https://doi.org/10.21468/SciPostPhys.11.2.045} {\bibfield  {journal}
  {\bibinfo  {journal} {SciPost Phys.}\ }\textbf {\bibinfo {volume} {11}},\
  \bibinfo {pages} {45} (\bibinfo {year} {2021})}\BibitemShut {NoStop}%
\bibitem [{\citenamefont {Tomasi}\ \emph {et~al.}(2019)\citenamefont {Tomasi},
  \citenamefont {Amini}, \citenamefont {Bera}, \citenamefont {Khaymovich},\
  and\ \citenamefont {Kravtsov}}]{Tomasi2019}%
  \BibitemOpen
  \bibfield  {author} {\bibinfo {author} {\bibfnamefont {G.~D.}\ \bibnamefont
  {Tomasi}}, \bibinfo {author} {\bibfnamefont {M.}~\bibnamefont {Amini}},
  \bibinfo {author} {\bibfnamefont {S.}~\bibnamefont {Bera}}, \bibinfo {author}
  {\bibfnamefont {I.~M.}\ \bibnamefont {Khaymovich}},\ and\ \bibinfo {author}
  {\bibfnamefont {V.~E.}\ \bibnamefont {Kravtsov}},\ }\bibfield  {title}
  {\bibinfo {title} {{Survival probability in Generalized {R}osenzweig-{P}orter
  random matrix ensemble}},\ }\href
  {https://doi.org/10.21468/SciPostPhys.6.1.014} {\bibfield  {journal}
  {\bibinfo  {journal} {SciPost Phys.}\ }\textbf {\bibinfo {volume} {6}},\
  \bibinfo {pages} {14} (\bibinfo {year} {2019})}\BibitemShut {NoStop}%
\bibitem [{\citenamefont {Bogomolny}\ and\ \citenamefont
  {Sieber}(2018)}]{Bogomolny2018}%
  \BibitemOpen
  \bibfield  {author} {\bibinfo {author} {\bibfnamefont {E.}~\bibnamefont
  {Bogomolny}}\ and\ \bibinfo {author} {\bibfnamefont {M.}~\bibnamefont
  {Sieber}},\ }\bibfield  {title} {\bibinfo {title} {Eigenfunction distribution
  for the {R}osenzweig-{P}orter model},\ }\href
  {https://doi.org/10.1103/PhysRevE.98.032139} {\bibfield  {journal} {\bibinfo
  {journal} {Phys. Rev. E}\ }\textbf {\bibinfo {volume} {98}},\ \bibinfo
  {pages} {032139} (\bibinfo {year} {2018})}\BibitemShut {NoStop}%
\bibitem [{\citenamefont {Facoetti}\ \emph {et~al.}(2016)\citenamefont
  {Facoetti}, \citenamefont {Vivo},\ and\ \citenamefont
  {Biroli}}]{Facoetti2016}%
  \BibitemOpen
  \bibfield  {author} {\bibinfo {author} {\bibfnamefont {D.}~\bibnamefont
  {Facoetti}}, \bibinfo {author} {\bibfnamefont {P.}~\bibnamefont {Vivo}},\
  and\ \bibinfo {author} {\bibfnamefont {G.}~\bibnamefont {Biroli}},\
  }\bibfield  {title} {\bibinfo {title} {From non-ergodic eigenvectors to local
  resolvent statistics and back: A random matrix perspective},\ }\href
  {https://doi.org/10.1209/0295-5075/115/47003} {\bibfield  {journal} {\bibinfo
   {journal} {{EPL} (Europhysics Letters)}\ }\textbf {\bibinfo {volume}
  {115}},\ \bibinfo {pages} {47003} (\bibinfo {year} {2016})}\BibitemShut
  {NoStop}%
\bibitem [{\citenamefont {Venturelli}\ \emph {et~al.}(2023)\citenamefont
  {Venturelli}, \citenamefont {Cugliandolo}, \citenamefont {Schehr},\ and\
  \citenamefont {Tarzia}}]{Venturelli2023}%
  \BibitemOpen
  \bibfield  {author} {\bibinfo {author} {\bibfnamefont {D.}~\bibnamefont
  {Venturelli}}, \bibinfo {author} {\bibfnamefont {L.~F.}\ \bibnamefont
  {Cugliandolo}}, \bibinfo {author} {\bibfnamefont {G.}~\bibnamefont
  {Schehr}},\ and\ \bibinfo {author} {\bibfnamefont {M.}~\bibnamefont
  {Tarzia}},\ }\bibfield  {title} {\bibinfo {title} {Replica approach to the
  generalized {R}osenzweig-{P}orter model},\ }\href
  {https://doi.org/10.21468/scipostphys.14.5.110} {\bibfield  {journal}
  {\bibinfo  {journal} {SciPost Physics}\ }\textbf {\bibinfo {volume} {14}},\
  \bibinfo {pages} {110} (\bibinfo {year} {2023})}\BibitemShut {NoStop}%
\bibitem [{\citenamefont {Das}\ and\ \citenamefont {Ghosh}(2019)}]{Das2019}%
  \BibitemOpen
  \bibfield  {author} {\bibinfo {author} {\bibfnamefont {A.~K.}\ \bibnamefont
  {Das}}\ and\ \bibinfo {author} {\bibfnamefont {A.}~\bibnamefont {Ghosh}},\
  }\bibfield  {title} {\bibinfo {title} {Eigenvalue statistics for generalized
  symmetric and {H}ermitian matrices},\ }\href
  {https://doi.org/10.1088/1751-8121/ab3711} {\bibfield  {journal} {\bibinfo
  {journal} {J. Phys. A}\ }\textbf {\bibinfo {volume} {52}},\ \bibinfo {pages}
  {395001} (\bibinfo {year} {2019})}\BibitemShut {NoStop}%
\bibitem [{\citenamefont {Santos}\ \emph {et~al.}(2004)\citenamefont {Santos},
  \citenamefont {Rigolin},\ and\ \citenamefont {Escobar}}]{SantosEscobar2004}%
  \BibitemOpen
  \bibfield  {author} {\bibinfo {author} {\bibfnamefont {L.~F.}\ \bibnamefont
  {Santos}}, \bibinfo {author} {\bibfnamefont {G.}~\bibnamefont {Rigolin}},\
  and\ \bibinfo {author} {\bibfnamefont {C.~O.}\ \bibnamefont {Escobar}},\
  }\bibfield  {title} {\bibinfo {title} {Entanglement versus chaos in
  disordered spin systems},\ }\href
  {https://doi.org/10.1103/PhysRevA.69.042304} {\bibfield  {journal} {\bibinfo
  {journal} {Phys. Rev. A}\ }\textbf {\bibinfo {volume} {69}},\ \bibinfo
  {pages} {042304} (\bibinfo {year} {2004})}\BibitemShut {NoStop}%
\bibitem [{\citenamefont {Dukesz}\ \emph {et~al.}(2009)\citenamefont {Dukesz},
  \citenamefont {Zilbergerts},\ and\ \citenamefont {Santos}}]{Dukesz2009}%
  \BibitemOpen
  \bibfield  {author} {\bibinfo {author} {\bibfnamefont {F.}~\bibnamefont
  {Dukesz}}, \bibinfo {author} {\bibfnamefont {M.}~\bibnamefont
  {Zilbergerts}},\ and\ \bibinfo {author} {\bibfnamefont {L.~F.}\ \bibnamefont
  {Santos}},\ }\bibfield  {title} {\bibinfo {title} {Interplay between
  interaction and (un)correlated disorder in one-dimensional many-particle
  systems: delocalization and global entanglement},\ }\href
  {https://doi.org/10.1088/1367-2630/11/4/043026} {\bibfield  {journal}
  {\bibinfo  {journal} {New J. Phys.}\ }\textbf {\bibinfo {volume} {11}},\
  \bibinfo {pages} {043026} (\bibinfo {year} {2009})}\BibitemShut {NoStop}%
\bibitem [{\citenamefont {Pal}\ and\ \citenamefont {Huse}(2010)}]{Pal2010}%
  \BibitemOpen
  \bibfield  {author} {\bibinfo {author} {\bibfnamefont {A.}~\bibnamefont
  {Pal}}\ and\ \bibinfo {author} {\bibfnamefont {D.~A.}\ \bibnamefont {Huse}},\
  }\bibfield  {title} {\bibinfo {title} {Many-body localization phase
  transition},\ }\href {https://doi.org/10.1103/PhysRevB.82.174411} {\bibfield
  {journal} {\bibinfo  {journal} {Phys. Rev. B}\ }\textbf {\bibinfo {volume}
  {82}},\ \bibinfo {pages} {174411} (\bibinfo {year} {2010})}\BibitemShut
  {NoStop}%
\bibitem [{\citenamefont {Schreiber}\ \emph {et~al.}(2015)\citenamefont
  {Schreiber}, \citenamefont {Hodgman}, \citenamefont {Bordia}, \citenamefont
  {L\"uschen}, \citenamefont {Fischer}, \citenamefont {Vosk}, \citenamefont
  {Altman}, \citenamefont {Schneider},\ and\ \citenamefont
  {Bloch}}]{Schreiber2015}%
  \BibitemOpen
  \bibfield  {author} {\bibinfo {author} {\bibfnamefont {M.}~\bibnamefont
  {Schreiber}}, \bibinfo {author} {\bibfnamefont {S.~S.}\ \bibnamefont
  {Hodgman}}, \bibinfo {author} {\bibfnamefont {P.}~\bibnamefont {Bordia}},
  \bibinfo {author} {\bibfnamefont {H.~P.}\ \bibnamefont {L\"uschen}}, \bibinfo
  {author} {\bibfnamefont {M.~H.}\ \bibnamefont {Fischer}}, \bibinfo {author}
  {\bibfnamefont {R.}~\bibnamefont {Vosk}}, \bibinfo {author} {\bibfnamefont
  {E.}~\bibnamefont {Altman}}, \bibinfo {author} {\bibfnamefont
  {U.}~\bibnamefont {Schneider}},\ and\ \bibinfo {author} {\bibfnamefont
  {I.}~\bibnamefont {Bloch}},\ }\bibfield  {title} {\bibinfo {title}
  {Observation of many-body localization of interacting fermions in a
  quasirandom optical lattice},\ }\href
  {https://doi.org/10.1126/science.aaa7432} {\bibfield  {journal} {\bibinfo
  {journal} {Science}\ }\textbf {\bibinfo {volume} {349}},\ \bibinfo {pages}
  {842} (\bibinfo {year} {2015})}\BibitemShut {NoStop}%
\bibitem [{\citenamefont {Luitz}\ and\ \citenamefont {Lev}(2017)}]{Luitz2017}%
  \BibitemOpen
  \bibfield  {author} {\bibinfo {author} {\bibfnamefont {D.}~\bibnamefont
  {Luitz}}\ and\ \bibinfo {author} {\bibfnamefont {Y.~B.}\ \bibnamefont
  {Lev}},\ }\bibfield  {title} {\bibinfo {title} {The ergodic side of the
  many-body localization transition},\ }\href
  {https://doi.org/10.1002/andp.201600350} {\bibfield  {journal} {\bibinfo
  {journal} {Ann. Phys.(Berlin)}\ }\textbf {\bibinfo {volume} {529}},\ \bibinfo
  {pages} {1600350} (\bibinfo {year} {2017})}\BibitemShut {NoStop}%
\bibitem [{\citenamefont {Sierant}\ \emph {et~al.}(2024)\citenamefont
  {Sierant}, \citenamefont {Lewenstein}, \citenamefont {Scardicchio},
  \citenamefont {Vidmar},\ and\ \citenamefont {Zakrzewski}}]{Sierant2024ARXIV}%
  \BibitemOpen
  \bibfield  {author} {\bibinfo {author} {\bibfnamefont {P.}~\bibnamefont
  {Sierant}}, \bibinfo {author} {\bibfnamefont {M.}~\bibnamefont {Lewenstein}},
  \bibinfo {author} {\bibfnamefont {A.}~\bibnamefont {Scardicchio}}, \bibinfo
  {author} {\bibfnamefont {L.}~\bibnamefont {Vidmar}},\ and\ \bibinfo {author}
  {\bibfnamefont {J.}~\bibnamefont {Zakrzewski}},\ }\href@noop {} {\bibinfo
  {title} {Many-body localization in the age of classical computing}} (\bibinfo
  {year} {2024}),\ \Eprint {https://arxiv.org/abs/2403.07111} {arXiv:2403.07111
  [cond-mat.dis-nn]} \BibitemShut {NoStop}%
\bibitem [{\citenamefont {Santos}\ and\ \citenamefont
  {Rigol}(2010)}]{Santos2010PRE}%
  \BibitemOpen
  \bibfield  {author} {\bibinfo {author} {\bibfnamefont {L.~F.}\ \bibnamefont
  {Santos}}\ and\ \bibinfo {author} {\bibfnamefont {M.}~\bibnamefont {Rigol}},\
  }\bibfield  {title} {\bibinfo {title} {Onset of quantum chaos in
  one-dimensional bosonic and fermionic systems and its relation to
  thermalization},\ }\href {https://doi.org/10.1103/PhysRevE.81.036206}
  {\bibfield  {journal} {\bibinfo  {journal} {Phys. Rev. E}\ }\textbf {\bibinfo
  {volume} {81}},\ \bibinfo {pages} {036206} (\bibinfo {year}
  {2010})}\BibitemShut {NoStop}%
\bibitem [{\citenamefont {Torres-Herrera}\ and\ \citenamefont
  {Santos}(2013)}]{Torres2013}%
  \BibitemOpen
  \bibfield  {author} {\bibinfo {author} {\bibfnamefont {E.~J.}\ \bibnamefont
  {Torres-Herrera}}\ and\ \bibinfo {author} {\bibfnamefont {L.~F.}\
  \bibnamefont {Santos}},\ }\bibfield  {title} {\bibinfo {title} {Effects of
  the interplay between initial state and {H}amiltonian on the thermalization
  of isolated quantum many-body systems},\ }\href
  {https://doi.org/10.1103/PhysRevE.88.042121} {\bibfield  {journal} {\bibinfo
  {journal} {Phys. Rev. E}\ }\textbf {\bibinfo {volume} {88}},\ \bibinfo
  {pages} {042121} (\bibinfo {year} {2013})}\BibitemShut {NoStop}%
\bibitem [{\citenamefont {French}\ and\ \citenamefont
  {Wong}(1970)}]{French1970}%
  \BibitemOpen
  \bibfield  {author} {\bibinfo {author} {\bibfnamefont {J.~B.}\ \bibnamefont
  {French}}\ and\ \bibinfo {author} {\bibfnamefont {S.~S.~M.}\ \bibnamefont
  {Wong}},\ }\bibfield  {title} {\bibinfo {title} {Validity of random matrix
  theories for many-particle systems},\ }\href
  {https://doi.org/https://doi.org/10.1016/0370-2693(70)90213-3} {\bibfield
  {journal} {\bibinfo  {journal} {Phys. Lett. B}\ }\textbf {\bibinfo {volume}
  {33}},\ \bibinfo {pages} {449} (\bibinfo {year} {1970})}\BibitemShut
  {NoStop}%
\end{thebibliography}

%

\end{document}